\documentclass[aip,reprint,tightenlines,floatfix]{revtex4-2}  %>>> use for US letter paper
\usepackage[]{graphicx}
\usepackage{dcolumn}%
\usepackage{bm}%

\let\baraccent=\= % rename builtin command \= to \baraccent
\renewcommand{\=}[1]{\stackrel{#1}{=}} % for putting nums above =

\begin{document} 
\title{Hydrogen incorporation into amorphous indium gallium zinc oxide thin-film transistors } 

\author{George W. Mattson$^1$, Kyle T. Vogt$^1$, John F. Wager$^2$ and Matt W. Graham$^1$}
\affiliation{Department of Physics, Oregon State University, Corvallis, OR 97331-6507, USA} 
\affiliation{School of EECS, Oregon State University, Corvallis, OR 97331-5501, USA}

%>>>> uncomment following for page numbers
% \pagestyle{plain}    
%>>>> uncomment following to start page numbering at 301 
%\setcounter{page}{301} 

\begin{abstract}
\indent	
%Optically transparent semiconductors like amorphous InGaZnO$_x$ (a-IGZO) often contain donor-like and acceptor-like traps in the subgap that can significantly impact device operation characteristics such as hysteresis and turn-on voltage. Hydrogen is thermally released from a passivation layer and its diffusivity-reactivity creates negative-U defects at interstitial a-IGZO sites. These hydrogen defects are electron donors seen by increasing negative turn-on voltage shifts in the device drain current-gate voltage ($\mathrm{I_{D} - V_{G}}$) transfer curves as hydrogen is released in top-gated a-IGZO thin film transistors (TFT). The hydrogen contribution to the subgap trap state density of states is revealed by measuring the ultrabroadband photoconduction (UBPC) of TFT across photon energies to within $\mathrm{0.3}$ eV of the conduction band minimum. Consistent with hydrogen acting as a negative-U defect, a broad peak within $\mathrm{0.4}$ eV of the valence band (VB) maximum gives a differential trap density that matches the hydrogen concentrations predicted by the $\mathrm{I_{D} - V_{G}}$ curve turn-on voltage shifts.Temperature-dependent and transient photoconduction trap lifetime data indicate hydrogen-induced neutralization of Zn vacancy ($\mathrm{V_{Zn}}$) states and $\mathrm{{[{H^+} - {O_{O}^{2-}}]}^{1-}}$ state formation.%
Within the subgap of amorphous oxide semiconductors like amorphous indium gallium zinc oxide (a-IGZO) are donor-like and acceptor-like states that control the operational physics of optically transparent thin-film transistors (TFTs).  Hydrogen incorporation into the channel layer of a top-gate a-IGZO TFT exists as an electron donor that causes an observed negative shift in the drain current-gate voltage ($\mathrm{I_{D} - V_{G}}$) transfer curve turn-on voltage.  Normally, hydrogen is thought to create shallow electronic states just below the conduction band mobility edge, with the donor ionization state controlled by equilibrium thermodynamics involving the position of the Fermi level with respect to the donor ionization energy. However, hydrogen does not behave as a normal donor as revealed by the subgap density of states (DoS) measured by the photoconduction response of top-gate a-IGZO TFTs to within 0.3 eV of the CBM edge. Specifically, the DoS shows a subgap peak above the valence band mobility edge growing at the same rate that $\mathrm{I_{D} - V_{G}}$ transfer curve measurements suggest that hydrogen was  incorporated into the channel layer.  Such hydrogen donor behavior in a-IGZO is anomalous and can be understood as follows: Non-bonded hydrogen ionization precedes its incorporation into the a-IGZO network as a bonded species. Ionized hydrogen bonds to a charged oxygen-on-an-oxygen-site anion, resulting in the formation of a defect complex denoted herein as, $\mathrm{{[{O_{O}^{2-}}{H^+}]}^{1-}}$. Formation of an $\mathrm{{[{O_{O}^{2-}}{H^+}]}^{1-}}$ defect complex creates a spectrally-broad ($\sim$0.3 eV FWHM) distribution of electronic states observed in the bandgap centered at $0.4$ eV above the valence band mobility edge.
\\

\end{abstract}
\keywords{amorphous IGZO, thin-film transistor, density of states, ultrabroadband photoconduction,  hydrogen defects, negative-U}

 \maketitle

%%%%%%%%%%%%%%%%%%%%%
\section{INTRODUCTION}

%Amorphous oxide semiconductors have emerged over the previous two decades as a leading material for the conducting channel of thin film transistors (TFTs) in flat-panel display technology. In particular, InGaZnO (a-IGZO) has proliferated in TFTs in recent years due to properties such as its room-temperature processibility, high flexibility and scalability, and low power consumption.\cite{KnNomura2004, Kamiya2010, Kwon2011} While a-IGZO offers significant advantages over competitors such as amorphous Si and low temperature polycrystalline silicon in terms of carrier mobility and processing cost, respectively,\cite{Wager2020} there remain non-trivial challenges related to its use in TFTs, particularly phenomena such as bias illumination stressing, in which illumination under bias causes persistent alteration of the device drain current-gate voltage ($\mathrm{I_{D} - V_{G}}$) transfer curve. Previous works have extensively explored these phenomena, and have linked them to the presence of a large number of trap states in the a-IGZO subgap that emerge due to its disordered structure.\cite{Jang2015, Kim2018, DeJamblinneDeMeux2018} Measuring and understanding the nature of the subgap states is therefore of fundamental importance to address device stability concerns in a-IGZO.

Amorphous indium gallium zinc oxide (a-IGZO) thin-film transistors (TFTs) constitute a well-established, maturing technology for the realization of commercial flat-panel display backplanes.\cite{KnNomura2004, Kamiya2010, Kwon2011,Wager2020} The performance of an a-IGZO TFT depends, to a large extent, on the subgap electronic state properties of the a-IGZO channel layer. Four primary types of intrinsic subgap states exist in a-IGZO, i.e., conduction band tail states (acceptor states exponentially distributed below the conduction band mobility edge), valence band tail states (donor states exponentially distributed above the valence band mobility edge), oxygen vacancy states (donor states forming a series of Gaussian-like peaks in the upper portion of the a-IGZO band gap), and metal vacancy states (acceptor states giving Gaussian-like peaks in the lower portion of the a-IGZO band gap).\cite{Vogt, Wager2017}. Both types of band tail states and oxygen vacancy states are ubiquitous in a-IGZO, while ultrabroadband photoconduction (UBPC) assessment reveals that the concentration of metal vacancy states is quite variable, e.g., $\mathrm{~10^{16}-10^{17}}$ cm$^{-3}$ for the bottom-gate a-IGZO TFTs of Vogt et al.\cite{Vogt} and negligibly small (i.e., $\mathrm{< 10^{16} cm^{-3}}$) for the top-gate a-IGZO TFTs examined herein.

In addition to \textit{intrinsic} subgap states, one type of \textit{extrinsic} subgap state is commonly present in a-IGZO, i.e., hydrogen. Although there is abundant literature on the topic of hydrogen incorporation into a-IGZO,\cite{Lee2009, Miyase2014, Bang2017, Li2017, Song2017, Han2017, Abliz2017, Nam2018, Felizco2020, Chen2020, Wang2020, Abliz2020, Corsino2020, Prasad2021, Magari2021} this literature is confusing and often contradictory or even erroneous. The concept of hydrogen interacting with a-IGZO as a donor, contributing an electron to the conduction band, is well established through both computational\cite{Li2017, Bang2017, DeJamblinneDeMeux2018} and experimental methods, the latter of which have commonly found a negative shift in the TFT $\mathrm{I_{D} - V_{G}}$ transfer curve turn-on voltage ($\mathrm{V_{ON}}$) after hydrogen incorporation.\cite{Lee2021, Chen2020, Jeong2020, Chen2020_hump, Liu2021, Kim2012, Kim2019, Song2021, Song2017}
\begin{figure*}
   \begin{center}
   \begin{tabular}{c}
   \includegraphics[height=6.5cm]{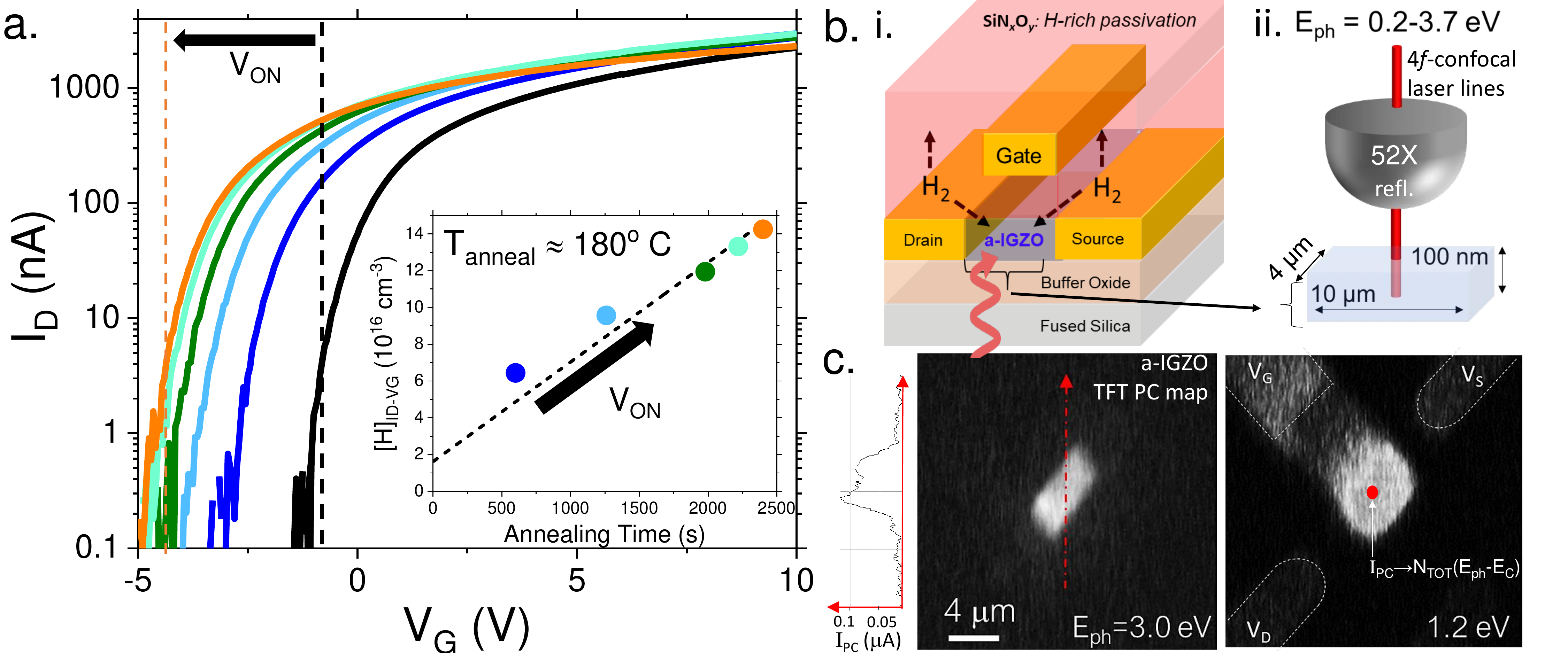}
   \end{tabular}
   \end{center}
   \caption[example] 
%>>>> use \label inside caption to get Fig. number with \ref{}
    {\label{fig1} 
\textbf{ \textbf{(a)}} Thermal annealing of a top-gate a-IGZO TFT leads to a negative shift in the drain current-gate voltage ($\mathrm{I_{D} - V_{G}}$) transfer curve turn-on voltage ($\mathrm{V_{ON}}$).  (\textit{inset}) Hydrogen incorporated into a-IGZO as estimated from $\mathrm{I_{D} - V_{G}}$ transfer curves ($\mathrm{[H]_{ID-VG}}$) as a function of annealing time. \textbf{(b) i.} Cross-sectional view of a typical top-gate a-IGZO TFT measured. When the TFT is thermally annealed at $\mathrm{180^{\circ} C}$, diffusion from the hydrogen-rich $\mathrm{SiN_{x}O_{y}}$ passivation layer dopes the adjacent a-IGZO active region with hydrogen.  \textbf{ii.} The ultrabroadband photoconduction (UBPC) method shown can measure how hydrogen diffusion impacts the total trap density, N$\mathrm{_{TOT}}$ by collecting photoconduction current (I$\mathrm{_{PC}}$) for photon excitation energies scanned from E$\mathrm{_{ph}}$ 0.3 to 3.7 eV.   \textbf{(c)} UBPC maps of the collected I$\mathrm{_{PC}}$ (see vertical PC-line cut) as a laser raster-scans over a top-gate TFT. Each map results from excitation of a-IGZO subgap traps above E$\mathrm{_{ph}}$-E$_C$= -3.0 or -1.2 eV. N$\mathrm{_{TOT}}$ and DoS spectra can then obtained from the diffraction-limited illumination centered in the active channel (\textit{white arrow}).}
   \end{figure*}

Hydrogen concentrations of order $\mathrm{10^{20} - 10^{21}}$ $\mathrm{cm^{-3}}$ are often observed in a-IGZO via secondary ion mass spectroscopy (SIMS) assessment;\cite{Bang2017} however, this hydrogen must be predominantly electrically inactive, or it would result in far larger negative shifts in the value of $\mathrm{V_{ON}}$ then are observed. It is generally agreed upon that the donor effect observed with hydrogen incorporation involves hydrogen bonding with oxygen atoms in the disordered a-IGZO lattice, yielding $\mathrm{{OH}^{-}}$.\cite{Kumomi2009, Nomura2013, Bang2017, DeJamblinneDeMeux2018, Song2021} Less robustly established, however, is the energetic location at which the resulting bond state manifests in the a-IGZO subgap. Density functional theory-based calculations have indicated a possible shallow donor state near ($\mathrm{<0.6}$ eV) from the conduction band minimum\cite{DeJamblinneDeMeux2018} and enhanced state density in hydrogen-rich devices has been reported to be observed in the range $\mathrm{0.2-0.6}$ eV from the conduction band minimum.\cite{Jang2021, Han2017} However, hydrogen incorporation in a-IGZO has also been repeatedly correlated with an increased response from states in the near valence band region of the subgap,\cite{Bang2017, Kang2018, Kim2019, Lee2021, Wang2020, Song2017} which at least one author\cite{Kang2018} has suggested could be linked to $\mathrm{OH^{-}}$-interaction.

More commonly, however, both observed changes in state density near the valence band and various electrical effects such as worsened bias illumination stressing in hydrogen-incorporated devices have been associated with the formation of a hydrogen-oxygen vacancy (or hydrogen-metal) complex.\cite{Bang2017, Li2017, DeJamblinneDeMeux2018} However, hydrogen passivation of oxygen vacancies cannot explain the negative transfer curve $\mathrm{V_{ON}}$ shift because the corresponding chemical reaction, i.e. $\mathrm{{H^0} + {V_{O}}^{0} \rightarrow {[HV_{O}]}^{0}}$ is electrically neutral.

There are many reports in the literature in which x-ray photoelectron spectroscopy (XPS) is (assertedly) employed for the assessment of oxygen vacancies and/or hydrogen donors present in bulk a-IGZO. Conclusions from these studies are not viable. Since the atomic sensitivity of XPS is $\mathrm{\sim 0.1-1}$ atomic percent and the atomic density of a-IGZO is $\mathrm{8.2 \times 10^{22}}$ $\mathrm{cm^{-3}}$, the best-case XPS sensitivity for the assessment of a-IGZO is $\mathrm{\sim 10^{20}}$ $\mathrm{cm^{-3}}$. However, oxygen vacancy or hydrogen donor concentrations in a-IGZO are typically on the order of $\mathrm{\sim 10^{15}-10^{18}}$ $\mathrm{cm^{-3}}$, multiple orders of magnitude below the detection limit of XPS. The pitfall associated with these erroneous literature claims is that these researchers have deconvolved small oxygen 1s XPS peak features associated with adventitious hydrocarbons or other adsorbed surface species, and interpreted (incorrectly) that these surface features correspond to bulk oxygen vacancy and/or hydroxide concentrations.\cite{Rajachidambaram2012, Du2014}

The objective of the work presented herein is to formulate a physical picture of how hydrogen incorporation is accomplished in a-IGZO and how its incorporation affects the electrical properties of an a-IGZO TFT. Electrical and electro-optic characterization establishes that increasing the hydrogen concentration in a-IGZO (i) increases the density of free electrons in the conduction band, thereby shifting the turn-on voltage of the a-IGZO TFT to lower voltages, and (ii) increases the density of subgap states centered at $\mathrm{0.4}$ eV above the valence band mobility edge and also increases the valence band tail state characteristic (Urbach\cite{Urbach1953}) energy. It is argued that hydrogen acts as an anomalous, non-equilibrium donor in a-IGZO as a consequence of its negative-U property in which the formation of a netural, non-bonded hydrogen is energetically unfavorable.

\section{EXPERIMENTAL METHODS}

\subsection{Hydrogen incorporation into a-IGZO top-gate TFTs}

To better measure the fundamental properties of hydrogen incorporation in this amorphous semiconductor, a-IGZO TFTs were fabricated in top-gate configuration. Figure 1(bi) shows a cross-sectional view of a top-gate TFT, and highlights that the H-rich $\mathrm{SiN_{x}O_{y}}$ passivation  layer is adjacent to the a-IGZO layer active region. Upon controlled thermally annealing, Fig. 1(bi) shows that the top-gate configuration enables facile hydrogen diffusion from the H-rich $\mathrm{SiN_{x}O_{y}}$ regions directly to the active channel of a-IGZO. Various top-gate TFTs were measured with channel widths ranging from $\mathrm{2.5-8}$ $\mathrm{\mu m}$ and channel lengths between $\mathrm{12-16}$ $\mathrm{\mu m}$. Unless otherwise noted,  measurements presented correspond to TFTs with a 100 nm a-IGZO layer and with channel dimensions shown in Fig. 1(bii).  The top-gate insulator capacitance density is 34.5 nFcm$^{-2}$.

To promote the diffusion of hydrogen from the H-rich passivation layer into the adjacent a-IGZO active channel, the TFT was thermally annealed at $\mathrm{180^{\circ} C}$, leading to a negative shift in the a-IGZO TFT turn-on voltage, $\mathrm{V_{ON}}$, with increasing annealing time. The inferred hydrogen concentration corresponding to the observed negative shift in $\mathrm{V_{ON}}$ is estimated as:

%for around one hour. During the thermal annealing process, Fig. 1(a) shows collected $\mathrm{I_{D} - V_{G}}$ curves continually shift toward more  negative voltages. As  annealing time increases,  the a-IGZO TFT threshold turn-on voltage, V$\mathrm{_{ON}}$ rigidly shifts negative.  

%Providing the negative $\mathrm{I_{D} - V_{G}}$ curves shifts are caused by the heat-induced hydrogen diffusion, V$\mathrm{_{ON}}$ can be used to estimate hydrogen trap density in the active channel. Previous works have suggested that molecular hydrogen dissociates into neutral hydrogen ($\mathrm{H^{o}}$) and bonds with a-IGZO.\cite{Domen2014} Assuming the reacted neutral hydrogen, $\mathrm{H^{0}}$ donates an electron to a-IGZO, the negative $\mathrm{V_{ON}}$ estimates hydrogen density, $\mathrm{[H]_{ID-VG}}$ as:

\begin{equation}
\mathrm{[H]_{ID-VG}} = \mathrm{ (-C_IV_{ON}/q)^{3/2}}                 
\end{equation} where $\mathrm{C_{I} = 34.5}$ $\mathrm{nF cm^{-2}}$ is the top-gate insulator capacitance density and q is electronic charge. Using Eq. 1, the inset of Fig. 1(a) plots the calculated $\mathrm{[H]_{ID-VG}}$ as a function of annealing time.  

\subsection{Ultrabroadband photoconduction and DoS}

To better understand how a negative shift in the $\mathrm{I_{D} - V_{G}}$ curve of an a-IZGO TFT is linked to hydrogen diffusion, the integrated sub-gap trap density ($\mathrm{N_{TOT}}$) is measured.  Direct observation of the subgap trap density has proved difficult in the past,\cite{Jang2015, Jia2018} because the relatively small concentration of traps in the subgap necessitates a roughly $\mathrm{10^{-6}}$ level of precision in order to resolve against background. Ultrabroadband photoconduction (UBPC) is a on-chip microscopy method using highly-tunable lasers to resolve the density of subgap traps through precision measurements of the TFT photoconductivity (PC). Figure 1(bi) shows the basic elements of UBPC with further details provided in the Supplemental Materials.  

 %A series of laser sources continuously tunable from approximately $\mathrm{0.3-3.7}$ eV illuminates a diffraction-limited region of an a-IGZO TFT operating under forward bias of approximately $\mathrm{5}$ V past $\mathrm{V_{ON}}$, which produces a contribution to $\mathrm{I_{D}}$ from photoexcitation of trapped carriers in the a-IGZO subgap.
 
 Spanning the full IR to UV range, measuring the subgap a-IGZO states uses multiple lasers continuously tunable from 0.3 to 3.7 eV. Specifically, a Coherent Chameleon Ti:Sapphire laser is coupled to an APE Compact optical parametric oscillator and a HarmoniXX second harmonic generation system.    To minimize spectral aberrations, we use all-reflective optics, including a $52X$ high-NA reflective objective. A piezo scanning mirror within a 4f confocal system couples the laser lines into an Olympus BX61W microscope with a homebuilt transport setup.  While the photon energy E$_{ph}$ is scanned step-wise from 0.3 eV to 3.7 eV, inverted RF source-drain electrical probes collect the PC induced by the laser (I$\mathrm{_{PC}}$).   Selected energies are repeated afterwards to ensure curves are reproducible.  Select UBPC curves were verified on a second UBPC microscope that instead scans a white-light supercontinuum laser (See Supplementary Materials).  I$_{PC}$ is detected by an current pre-amplifier and a lock-in amplifier (Zurich HFLI) referenced to an optical chopper frequency modulating the laser at $585$ Hz.  Dark-curve background and illuminated transfer curve hysteric drift is removed from the I$\mathrm{_{PC}}$ signal by careful selection of the lock-in amplifier laser-chopping frequency. 
 
Figure 1(c) shows PC-maps generated by raster-scanning the laser at E$\mathrm{_{ph}}$ over a top-gate a-IGZO TFT. It produces a spatially uniform I$\mathrm{_{PC}}$ response over the a-IGZO TFT active region. Precision scanning optics maintain a diffraction-limited spot at the  a-IGZO TFT center for the duration of UBPC measurement. The I$\mathrm{_{PC}}$ signal must be normalized by the number of photons incident per second on the active region, $\mathrm{N_{ph}}$.  To accomplish this, both the laser power transmitted through the objective and through-objective+TFT chip are measured at each photon energy. The resulting measured signal, qI$\mathrm{_{PC}}$/$\mathrm{N_{ph}}$ is directly proportional to the total integrated trap density, N$\mathrm{_{TOT}}$ from the conduction band mobility edge to the photon energy, E$\mathrm{_{ph}}$ according to:

\begin{equation}
\mathrm{N_{TOT}(E_{ph})} = \left( \mathrm{\frac{I_{PC}}{N_{ph}}}\right) \left(  \mathrm{\frac{qC_{I}}{m}} \right)  \mathrm{\frac{N_{o,max}}{d}}
\end{equation}  where d is the thickness of a-IGZO,  and m is the slope of the dark $\mathrm{I_{D} - V_{G}}$ transfer curve in the linear region within $\mathrm{\pm 0.5}$ V of the constant 5 V forward gate bias voltage over which which the UBPC signal, I$\mathrm{_{PC}}$ was collected.  $\mathrm{N_{o,max}}$ is a constant calibration factor obtained by finding the maximum number of incident photons per second that give a measurable increase in the PC-response, i.e., the saturation photon flux measured at the near-band gap. Note that the notation  $\mathrm{E-E_C= -E_{ph}}$ is used interchangeably. 

Finally, the a-IGZO experimental density of states (DoS) is obtained by taking a derivative of the total subgap trap density, or $\mathrm{DoS(E_{ph})}=\frac{d\mathrm{N_{TOT}}}{d\mathrm{E_{ph}}} $.  For the as-grown a-IGZO TFTs, the experimental DoS peaks were simulated as gaussian peaks and exponential Urbach tails. Then using the [H]$\mathrm{_{ID-VG}}$ values obtained from $\mathrm{I_{D} - V_{G}}$ transfer curve measurements and Tauc band-gap analysis (see see Supplementary Materials), a simulated hydrogen-dependent DoS is compared again with the experimental curve.

%(shown in the scanning photoconduction maps in Fig. 1(d)) allows for direct experimental measurements of the a-IGZO subgap trap density.\cite{Vogt}  A series of laser sources continuously tunable from approximately $\mathrm{0.3-3.7}$ eV illuminates a diffraction-limited region of an a-IGZO TFT operating under forward bias of approximately $\mathrm{5}$ V past $\mathrm{V_{ON}}$, which produces a contribution to $\mathrm{I_{D}}$ from photoexcitation of trapped carriers in the a-IGZO subgap. As shown in Fig. 1(b), the a-IGZO TFT is illuminated with a nearly diffraction-limited laser spot at each photon energy, to yield a spatially uniform photoconduction map over the TFT active region. This photoconduction signal is collected through optically-chopped ($\mathrm{585}$ Hz) lock-in amplification which enables $\mathrm{> 10^6}$ signal to noise and largely eliminates illuminated hysteric drift contribution. The resulting photoconduction signal is proportional to the integrated trap density from the CBM to the subgap photon energy selected. Additional details of the UBPC experiment are provided in the supplement.

\section{EXPERIMENTAL RESULTS AND DISCUSSION}

\begin{figure*}
   \begin{center}
   \begin{tabular}{c}
   \includegraphics[height=16cm]{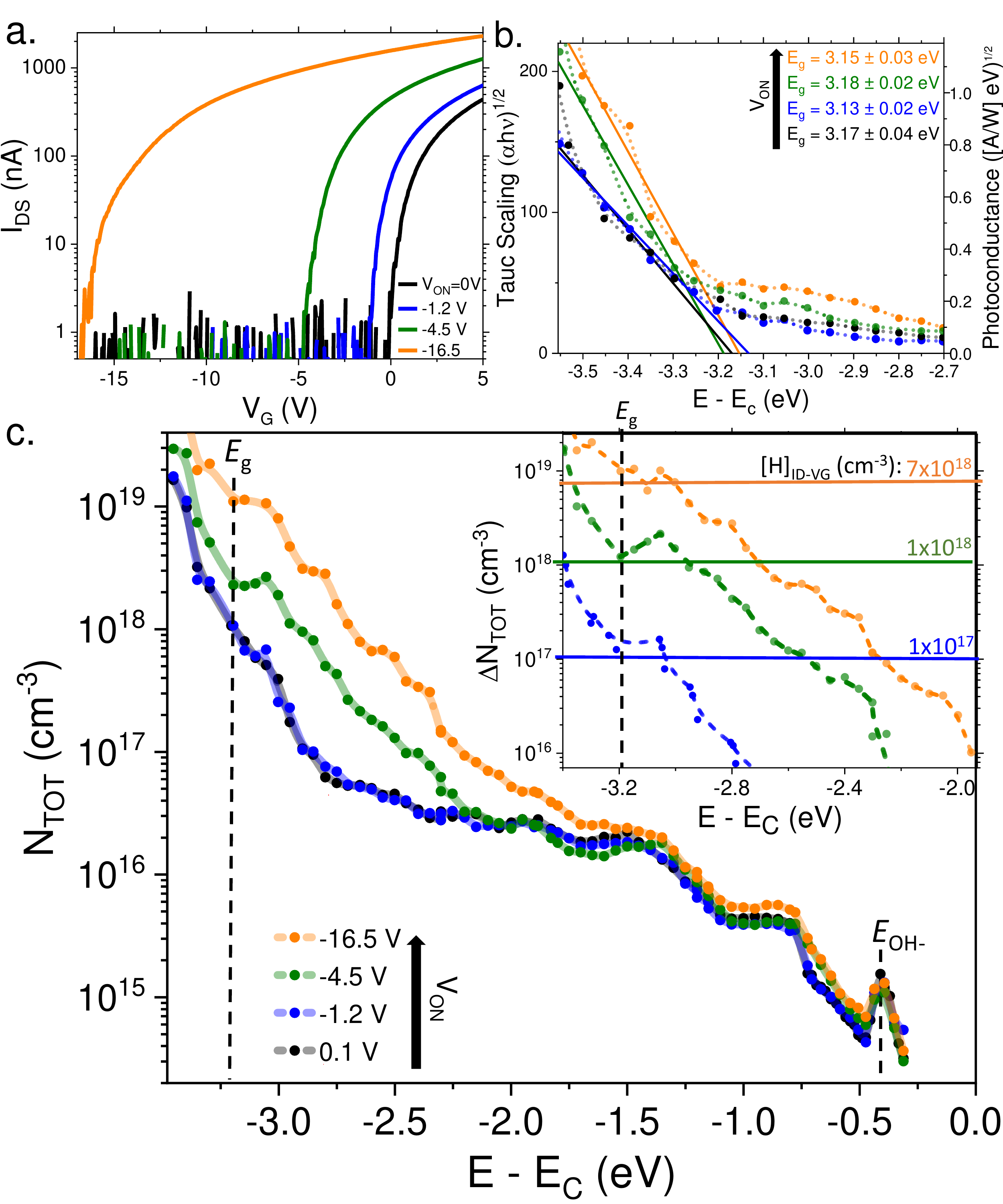}
   \end{tabular}
   \end{center}
   \caption[example] 
%>>>> use \label inside caption to get Fig. number with \ref{}
   { \label{fig2} 
\textbf{(a)} Drain current-gate voltage ($\mathrm{I_{D} - V_{G}}$) transfer curves for four top-gate a-IGZO TFTs with varying levels of hydrogen incorporation; a larger hydrogen concentration in the a-IGZO channel layer corresponds to a more negative shift in the turn-on voltage. \textbf{(b)} The  UBPC photoconduction signal estimates the band-gap of top-gate a-IGZO TFTs.  For four different $\mathrm{V_{ON}}$ device conditions, the Tauc plot scaling linear fits are shown (solid lines).  The average band-gap is $\mathrm{\sim3.17}$ eV.
\textbf{(c)} The UBPC total integrated trap density, $\mathrm{N_{TOT}}$, is plotted from $\mathrm{0.3}$ eV below the conduction band to the valence band region. (\textit{inset}) Plot of the differential or background-subtracted ($\mathrm{V_{ON} = 0}$ V curve) trap density maximizes the contribution from subgap states that contribute to each observed $\mathrm{I_{D} - V_{G}}$ $\mathrm{V_{ON}}$ shift. The  colored dashed lines correspond to the hydrogen concentration calculated directly from $\mathrm{I_{D} - V_{G}}$ transfer curves. At the valence band mobility edge ($\mathrm{E_G}$, vertical line), these $\mathrm{I_{D} - V_{G}}$ estimates are similar to the UBPC trap density measurement  value.}
   \end{figure*} 

Figure 2(a) shows $\mathrm{I_{D} - V_{G}}$ curves for four top-gate a-IGZO TFTs possessing varying concentrations of hydrogen (see Section II(A) for an explanation of how hydrogen is introduced into the a-IGZO TFT channel layer). A higher concentration of hydrogen corresponds to a more negative turn-on voltage, $\mathrm{V_{ON}}$, indicating that incorporated hydrogen is electrically active and that hydrogen behaves as a donor dopant since it increases the density of free conduction band electrons within the a-IGZO channel layer, thereby shifting $\mathrm{V_{ON}}$ to a more negative voltage.

Figure 2(b) presents Tauc photoconductance plots for four top-gate a-IGZO TFTs possessing varying concentrations of hydrogen. Details of the Tauc scaling of the UBPC photoconduction data are provided in the Supplementary Materials. The optical band gap is estimated via linear regression fitting of the valence band onset portion of the Tauc plot, yielding an average bandgap of $\mathrm{E_{G} = 3.15 \pm 0.03}$ eV for a-IGZO. While the a-IGZO band gap varies only slightly with increasing hydrogen incorporation, the Tauc plot in the subgap region  increases with increasing hydrogen concentration for subgap states positioned within $\mathrm{\sim 0.4}$ eV of the valence band mobility edge. Thus, Tauc photoconductance plots indicate that increasing hydrogen concentration increases the density of subgap electronic states located within $\mathrm{\sim 0.4}$ eV above the valence band mobility edge.

Figure 2(c) shows the total integrated trap density, $\mathrm{N_{TOT} ({cm}^{-3})}$, for four top-gate a-IGZO TFTs possessing varying concentrations of hydrogen. $\mathrm{N_{TOT}}$ is evaluated via ultrabroadband photoconduction (UBPC) as described in Section II(B). While $\mathrm{N_{TOT}}$ increases monotonically with increasing hydrogen concentration, this increase occurs almost exclusively in the near-valence band portion of the band gap. In the upper portions of the band gap, from $\sim$-2 eV to -0.5 eV,  the $\mathrm{N_{TOT}}$ signal originates from mainly the photoionization of oxygen vacancy related subgap traps, and is independent of the incorporated hydrogen concentration. Thus, the UBPC data demonstrates that hydrogen incorporation into a-IGZO gives rise to the creation of subgap electronic states located in the lower portion of the band gap.

The inset of Fig. 2(c) offers another perspective on hydrogen-induced subgap trap creation using UBPC trap density difference curves, obtained by taking  $\mathrm{\Delta N_{TOT}(V_{ON}) =}$ $\mathrm{N_{TOT}(V_{ON})- N_{TOT}(V_{ON} = 0}$ V). As before, an increase in trap density near the valence band mobility edge occurs as hydrogen incorporation increases. The logarithmic ordinate scale employed in Fig. 2(c) uses a seven orders of magnitude scale to explore the subtle density of states features associated with the high dynamic range of UBPC.  Comparison against the linear ordinate scaling shown in Supplemental Figure S4 is helpful to emphasize that hydrogen-induced subgap state creation primarily occurs within $\mathrm{\sim 0.4}$ eV of the valence band mobility edge.

\begin{center}
\begin{table}
 \begin{tabular}{c| c c c} 

 $\Delta \mathrm{V_{ON} (V)}$  & $\mathrm{[H]_{ID-VG} (cm^{-3})}$ & $\mathrm{[H]_{UBPC} (cm^{-3})}$ & $\Delta \mathrm{[H]} (cm^{-3})$ \\ [0.5ex] 
 \hline

$-1.2$  & $1 \times 10^{17}$ & $2.0 \times 10^{17}$  & $1 \times 10^{17}$\\

$-4.5$ & $1 \times 10^{18}$ & $1.6 \times 10^{18}$ & $6 \times 10^{17}$ \\

$-16.5$  & $7 \times 10^{18}$ & $1.0 \times 10^{19}$ & $3 \times 10^{18}$ \\ 

 \end{tabular}
  \caption[example]
 { \label{table1} 
Comparison of hydrogen concentration estimates from the four top-gate a-IGZO TFTs in Fig. 2.  Estimates obtained from $\mathrm{I_{D} - V_{G}}$ transfer curve shifts (Fig. 2a) are compared to measurements of the UBPC differential trap density taken at the valence band mobility edge or at  -3.17 eV in Fig. 2c (\textit{inset}). The $\Delta \mathrm{[H]}$ column shows that the UBPC estimates give $\sim30\%$ excess hydrogen compared to $\mathrm{I_{D} - V_{G}}$ estimates.
}
 \end{table}
\end{center}
 
 \indent How much hydrogen is incorporated into the four a-IGZO TFTs under consideration? Two methods are employed to estimate the incorporated hydrogen concentration. From $\mathrm{I_{D} - V_{G}}$ transfer curves (Fig. 2(a)), the incorporated hydrogen is estimated using $\mathrm{[H]_{ID-VG} = {(-{C_{I}} V_{ON} / q)}^{3/2}}$ where $\mathrm{C_{I} = 34.5 nF cm^{-2}}$ is the gate insulator capacitance density and q is electronic charge. From differential trap density measurement, the incorporated hydrogen concentration is estimated using $\mathrm{[H]_{UBPC} = \Delta N_{TOT} (E = E_{G} \cong -3.2)}$ eV. The inset of Fig. 2(c) shows the $\mathrm{[H]_{I_{D} - V_{G}}}$ (horizontal lines) are indeed very similar to the UBPC $\Delta \mathrm{N_{TOT}}$ measurements taken at intercept to the $E_G$ vertical dashed line. 

\begin{figure}
   \begin{center}
   \begin{tabular}{c}
   \includegraphics[height=14cm]{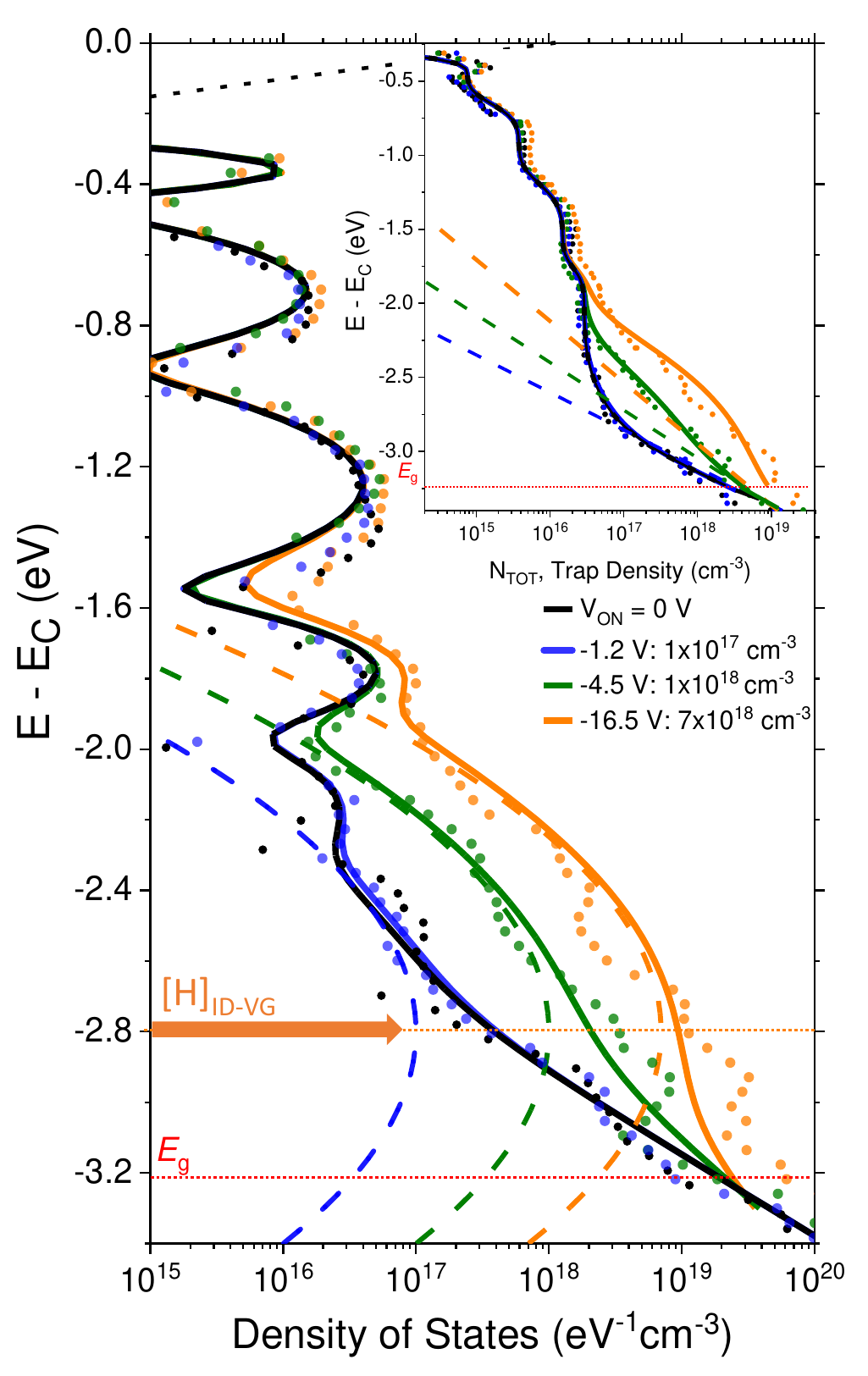}
   \end{tabular}
   \end{center}
   \caption[example] 
%>>>> use \label inside caption to get Fig. number with \ref{}
   { \label{fig3} 
Plot of  the experimental (\textit{closed circles}) subgap DoS of top-gate a-IGZO TFTs with varying degrees of hydrogen incorporation.  Overlaid is a simulation (\textit{solid lines}) of the convolved subgap defect peaks and valence band Urbach tail. The Gaussian peak centered (\textit{dashed lines}) at $\mathrm{-2.8}$ eV is most salient for the $\mathrm{V_{ON}= -4.5}$ and $\mathrm{-16.5}$ V shifted curves, and its simulated peak density was constrained to be equal to the $\mathrm{[H]_{ID-VG}}$ trap density.  After all peaks of the $\mathrm{V_{ON}= 0}$ V case were simulated, the Urbach tail energy was the only free simulation parameter ($\mathrm{95}$ - $\mathrm{180}$ meV). \textbf{(\textit{inset})} The corresponding experimental (\textit{closed circles}) and the integral of the simulated DoS  (\textit{solid lines}) confirms the DoS simulation predicts the UBPC trap density hydrogen dependence. \textit{Dashed lines} are the simulated change in valence band Urbach energy.}
\end{figure}

While Table I shows that the hydrogen concentrations estimated from $\mathrm{I_{D} - V_{G}}$ transfer curves and UBPC are similar, $\mathrm{[H]_{UBPC}}$ is always found to be larger than $\mathrm{[H]_{ID-VG}}$, by a factor of $\mathrm{\sim 2}$. This suggests that hydrogen incorporation in a-IGZO leads to the creation of two types of electronic subgap states. First, hydrogen donor states ($\mathrm{[H]_{ID-VG}}$) are electrically active and form a Gaussian-like band centered at $\mathrm{\sim 0.4}$ eV above the valence band mobility edge. Second, hydrogen-induced valence band tail states ($\mathrm{[H]_{UBPC}} - \mathrm{[H]_{ID-VG}}$) constitute an enhancement of the valence band tail state density (and hence the valence band Urbach energy, $\mathrm{W_{TD}}$) due to the incorporation of hydrogen (which increased disorder on the anion (oxygen) sublattice); since valence band tail states are donor-like, they are electrically neutral, and hence, electrically inactive from the perspective of a-IGZO TFT operation.

%These two estimates of the incorporated hydrogen concentration for the four a-IGZO TFTs under consideration are included in Table I. The agreement between $\mathrm{[H]_{ID-VG} \cong [H]_{UBPC}}$ means that the hydrogen-induced conduction band free electron concentration is approximately equal to the concentration of hydrogen-induced near valence band subgap states created, implying that all of the incorporated hydrogen measured via UBPC is electrically active.

Figure 3 shows a comparison between the experimental subgap density of states (DoS) of the four a-IGZO TFTs from the $\mathrm{N_{TOT}}$ UBPC measurement.  The solid lines are simulations based on a Gaussian hydrogen donor subgap peak centered at $\mathrm{-2.8}$ eV convolved with an Urbach\cite{Urbach1953,Wager2017} tail whose characteristic energy increases monotonically with hydrogen concentration due to increased disorder of the anion sublattice from which the valence band tail states originate. The simulation reproduces both the experimental density of states and trap density (shown in the inset of Fig. 3). Based on the simulation and experimental data, the composition of the a-IGZO subgap is quantified in Table II. The simulated peak maximum values and Urbach tail energies corresponding to the density of states spectra for the a-IGZO TFTs of varying hydrogen levels shown in Fig. 3 are provided in Table III.   The simulation response is dominated by the -2.8 eV peak whose density was fixed to that of the $\mathrm{[H]_{ID-VG}}$ values. As such, the Urbach energy, $\mathrm{E_U}$ in Table III, is the only free parameter required to match the simulation to the experiment. 

%$\mathrm{0 V}$  & $\mathrm{-1.2 V}$ & $\mathrm{-4.5 V}$ & $\mathrm{-16.5 V}$

\begin{center}
\begin{table}
 \begin{tabular}{c |c c c} 

 Peak & $\mathrm{E_{Peak}}$ (eV) & Max ($\mathrm{cm^{-3}eV^{-1}}$) & FWHM (eV) \\ [0.5ex]
 \hline
$\mathrm{{[{O_{O}^{2-}}{H^+}]}^{1-}}$ & $\mathrm{-2.8}$ & see Table III & $\mathrm{0.32}$ \\
$\mathrm{V_{M}}$ & $\mathrm{-2.15}$ & $2 \times 10^{15}$ & $\mathrm{0.10}$ \\

$\mathrm{V_{O}}$ & $\mathrm{-1.78}$ & $7 \times 10^{16}$ & $\mathrm{0.10}$ \\

$\mathrm{V_{O}}$ & $\mathrm{-1.25}$ & $4 \times 10^{16}$ & $\mathrm{0.13}$ \\
 
$\mathrm{V_{O}}$ & $\mathrm{-0.7}$ & $1.5 \times 10^{16}$ & $\mathrm{0.09}$ \\ 

 $\mathrm{OH_{Vibronic}}$ & $\mathrm{-0.37}$ & $1 \times 10^{16}$ & $\mathrm{0.05}$ \\ 
 
 \end{tabular}
 \caption[example]
 { \label{table2} 
Figures of merit extracted from simulation of the DoS spectra in Fig. 3. The first column identifies our suggested peak assignments as originating from $\mathrm{{[{O_{O}^{2-}}{H^+}]}^{1-}}$ bonds, metal vacancies V$_M$, oxygen vacancies V$_O$, or OH stretching mode. Subsequent columns contain the respective peak energies, peak trap densities, and  the FWHM spectral width of each peak.
}
\end{table}
\end{center}

\begin{center}
\begin{table}
 \begin{tabular}{c| c c} 
 
 $\mathrm{\Delta V_{ON}}$ (V) & $\mathrm{[H]_{ID-VG}}$ $\mathrm{(cm^{-3})}$ & $\mathrm{E_{U}}$ (meV) \\ [0.5ex]
 \hline
$\mathrm{0}$ & $1 \times 10^{16}$ & $\mathrm{95}$ \\

$\mathrm{-1.2}$ & $1 \times 10^{17}$ & $\mathrm{100}$ \\

$\mathrm{-4.5}$ & $1 \times 10^{18}$ & $\mathrm{140}$ \\ 

 $\mathrm{-16.5}$ & $7 \times 10^{18}$ & $\mathrm{180}$ \\ 

%>>>> use \label inside caption to get Fig. number with \ref{}
\end{tabular}
\caption[example]
   { \label{table3} 
Simulated density of states parameters for the spectra of a-IGZO TFTs with varying concentrations of hydrogen. Max DoS amplitudes are fixed according to the values determined from the $\mathrm{{I_D}-{V_G}}$ transfer curve shift. The $\mathrm{{[{O_{O}^{2-}}{H^+}]}^{1-}}$ peak energy is at at  $\mathrm{-2.8}$ eV or 0.4 eV above the VBM. $\mathrm{E_{U}}$ is the Urbach energy, the only free  parameter needed to simulate the experimental trap density and DoS spectra shown in Fig. 3. 
}
\end{table}
\end{center}

\section{DISCUSSION: Hydrogen in $\mathrm{a}$-IGZO}

$\mathrm{[H]_{ID-VG}}$ in Table I is an estimate of the electrically-active hydrogen incorporated into a-IGZO that gives rise to a corresponding increase in the subgap state density just above the valence band mobility edge. Since valence band states are derived primarily from O $\mathrm{2}$p atomic orbital basis states, it is clear that incorporated hydrogen mainly interacts with the oxygen anion sublattice, as described by the defect reaction

\begin{equation}
\mathrm{O_{O}^{2-}} + \mathrm{H^{0}} + \rightarrow \mathrm{{[ {O_{O}^{2-}}{H^+} ]}^{1-}} + \mathrm{e^{-}}
\end{equation} where $\mathrm{O_{O}^{2-}}$ denotes a bonded oxygen ion sitting on an oxygen site that is embedded within the a-IGZO amorphous network, presuming an ionic bonding model such that the oxygen ion is assumed to possess a $\mathrm{2-}$ charge state, $\mathrm{H^0}$ is a neutral hydrogen atom being incorporated into the a-IGZO network, $\mathrm{{[{O_{O}^{2-}}{H^+}]}^{1-}}$ is a defect complex consisting of a bonded $\mathrm{O_{O}^{2-}}$ oxygen interacting coulombically (primarily) with a positively ionized hydrogen atom such that an overall charge state of $\mathrm{1-}$ exists for this defect complex (it is possible to more simply and equivalently identify this defect complex as $\mathrm{OH^{-}}$ but we prefer to use the $\mathrm{{[{O_{O}^{2-}}{H^+}]}^{1-}}$ notation as it more accurately captures the detailed nature of the interaction (bonding) between oxygen and hydrogen), and $\mathrm{e^{-}}$ is a free conduction band electron giving rise to the negative turn-on voltage shift witnessed in Fig. 2(a). Note that the presence of free electrons on the right side of the Eq. (3) defect reaction is consistent with hydrogen behaving as a donor.

Isolating the hydrogen portion of the defect reaction discussed above, i.e., $\mathrm{H^0 \rightarrow H^{+} + e^{-}}$, demonstrates that hydrogen incorporation involves donor ionization, where hydrogen is the donor. This observation leads to the following question: How is it possible for donor ionization to occur from subgap states residing just above the valence band mobility edge? Clearly, donor ionization cannot be accomplished in the normal manner, i.e., by thermal emission of a neutral donor electron to the conduction band, since the H-induced subgap states are located $\mathrm{\sim 2.8}$ eV below the conduction band mobility edge. As discussed below, understanding donor ionization in the context of hydrogen incorporation requires elucidation of the negative-U character of hydrogen,\cite{VanDeWalle2003, VanDeWalle2006Review}.

%While fitting a single Urbach energy to each photocurrent spectrum gives a general idea of the hydrogen broadening of VB tail states, the high-H UBPC spectra also exhibit additional features consistent with an H-induced vibronic mode. For instance, both the $4.4$ V and $16.2$ V-shifted curves in Figure 1b show an enhanced photocurrent response within around $0.4$ eV from the device bandgap. Photoluminescence data from a $100$ nm a-IGZO film indicates the presence of this vibronic peak even in as-grown a-IGZO, and the higher-H samples show reduced transmission in the OH-stretch vibronic band near the CB minimum consistent with the introduction of OH states via hydrogen incorporation (See Supplement). These data suggest that the OH vibronic band approximately $0.4$ eV from the VB maximum is expanded in the high-H devices as a result of greater OH concentration. Laser heating from the UBPC illumination source vibronically excites VB carriers into this expanded sideband and, in concert with the energy carried by incident photons, provides sufficient energy for the carriers to exceed the band-gap. However, the predominant effect on the a-IGZO TFT trap density is the VBTS broadening quantified above, and the vibronic enhancement only causes significant divergence from linearity in the VBTS for the highest-H sample measured.
	
\begin{figure*}
   \begin{center}
   \begin{tabular}{c}
   \includegraphics[height=5cm]{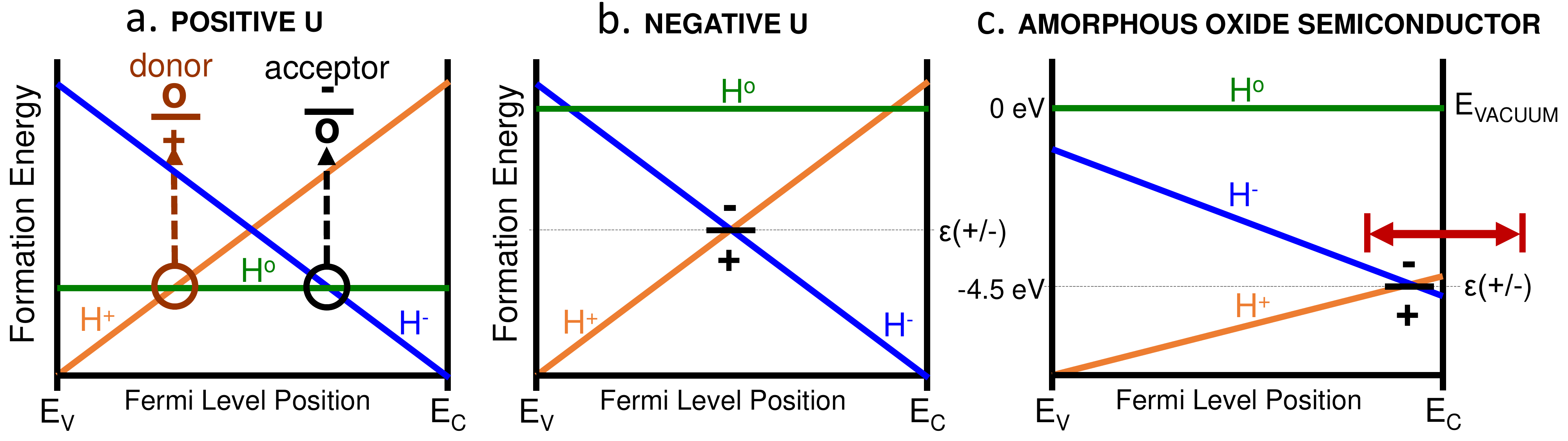}
   \end{tabular}
   \end{center}
   \caption[example] 
%>>>> use \label inside caption to get Fig. number with \ref{}
   { \label{fig5} 
\textbf{(a)} Positive U and \textbf{(b)} negative U formation energy versus Fermi level position trends for hydrogen incorporation into a generic semiconductor or insulator. In contrast, \textbf{(c)} displays the expected Fermi level position trend for hydrogen incorporation into an amorphous oxide semiconductor such as a-IGZO, with the ordinate corresponding to the energies referenced to the vacuum level.}
   \end{figure*} 

Figure 4(a) illustrates in a generic, idealized sense the formation energy as a function of Fermi level position trend for hydrogen residing within a semiconductor or insulator solid \textit{assuming} that hydrogen behaves as a normal or positive-U defect (which it does \textit{not}; U denotes correlation energy, as discussed below). Three charge states are possible for hydrogen, i.e., neutral ($\mathrm{H^{0}}$), positively ionized ($\mathrm{H^{+}}$), or negatively ionized ($\mathrm{H^{-}}$). The trend according to Fig. 4(a) is that $\mathrm{H^{+}}$ is energetically preferred when the Fermi level is located near the valence band, while $\mathrm{H^{-}}$ has the lowest formation energy when the Fermi level is close to the conduction band; $\mathrm{H^{0}}$ is energetically favorable when the Fermi level is near midgap. The type of defect behavior shown in Fig. 4(a) results in two distinct ionization energies in which (i) a positive charge state transitions to a neutral charge state (donor neutralization) and (ii) a neutral charge state transitions to a negative charge state (acceptor ionization) as the Fermi level is modulated across the bandgap from the valence band to the conduction band mobility edge. Notice that each charge state transition shown in Fig. 4(a) involves only one electron. This type of formation energy-Fermi level trend is often classified as `normal' since it is typically expected that a charge state transition involves only one electron.

In contrast, Fig. 4(b) exhibits a different type of formation energy as a function of Fermi level position trend as a consequence of a re-positioning of the neutral $\mathrm{H^{0}}$ formation energy (which does not depend on the Fermi level) to a higher energy than that shown in Fig. 4(a). This leads to a situation in which only one ionization energy exists, and any charge state transition involves two electrons. For our purposes, negative-U behavior of hydrogen in a solid corresponds to a situation as shown in Fig. 4(b), in which the neutral $\mathrm{H^{0}}$  formation energy is always larger than that of the $\mathrm{H^{+}}$  or $\mathrm{H^{-}}$ formation energy, depending on where the Fermi level is positioned. Quantitatively, if we define correlation energy as $\mathrm{U = {E_{F}}(0/-) - {E_{F}}(+/0)}$ (other definitions for correlation energy are possible), then the positive- or negative-U nature of hydrogen in a solid is clear and unambiguous.

Figures 4(a) and 4(b) are generic, idealized examples of positive-U and negative-U trends for hydrogen in a semiconductor or insulator solid. Figure 4(c) is a more realistic version of what the formation energy-Fermi level position might look like for an amorphous oxide semiconductor such as a-IGZO. Three aspects of Fig. 4(c) are important. First, hydrogen is a negative-U defect in a-IGZO. Second, as indicated by the red arrow, the $\mathrm{+/-}$ ionization energy for a-IGZO is likely to be positioned near to or (more likely) degenerate with the conduction band, rather than in the middle of the band-gap as shown in Figs. 4(a) and 4(b). This means that the $\mathrm{H^{+}}$ charge state is more likely to exist in a-IGZO than the $\mathrm{H^{-}}$ charge state. Third, the ordinate of Fig. 4(c) has been changed to illustrate how it might look with respect to the vacuum level, in accordance with the universal alignment of hydrogen theory of Van de Walle and Neugebauer,\cite{VanDeWalle2003,VanDeWalle2006Review} while the neutral hydrogen formation energy is positioned at the vacuum level.

Energy positioning of the hydrogen $\mathrm{+/-}$ energy and $\mathrm{H^{0}}$ merits a bit more clarification. Van de Walle and Neugebauer assert that the hydrogen $\mathrm{+/-}$ ionization energy is invariably positioned at or near $\mathrm{-4.5}$ eV from the vacuum level for all semiconductors, insulators, and even for aqueous solutions, and that this universal positioning corresponds to the standard hydrogen electrode potential of electrochemistry.\cite{VanDeWalle2003,VanDeWalle2006Review} Note that the bond dissociation energy of molecular hydrogen, $\mathrm{H_{2}}$, is  $\mathrm{435.7}$ kJ/mol or $\mathrm{4.5}$ eV. Thus, when molecular hydrogen is present in a-IGZO, it is energetically positioned at $\mathrm{-4.5}$ eV below the vacuum level, corresponding to its potential or bond dissociation energy. If $\mathrm{H_{2}}$ dissociates into two neutral atoms, $\mathrm{2H^{0}}$, within the a-IGZO network, these neutral hydrogen atoms are neither bonded to the a-IGZO lattice nor to themselves such that they only possess a small amount of kinetic energy and are positioned slightly above the vacuum level, as shown in Fig. 4(c).

Finally, after this extended negative-U behavior detour, we are now in an excellent position to answer the previously posed question: How is it possible for donor ionization to occur from subgap states residing just above the valence band mobility edge? The answer is that hydrogen donors behave differently from other donors. Most donors are first embedded into the lattice network, after which donor ionization occurs. In contrast, donor ionization occurs when hydrogen is in its neutral charge state, prior to its incorporation into the lattice. After ionization, hydrogen is incorporated into the network as a positively charged species. Since $\mathrm{H^{0}}$ is positioned at or near the vacuum level, no donor ionization energy barrier exists for hydrogen and the energetic positioning of its subgap states is irrelevant.

\begin{figure}
   \begin{center}
   \begin{tabular}{c}
   \includegraphics[height=8.5cm]{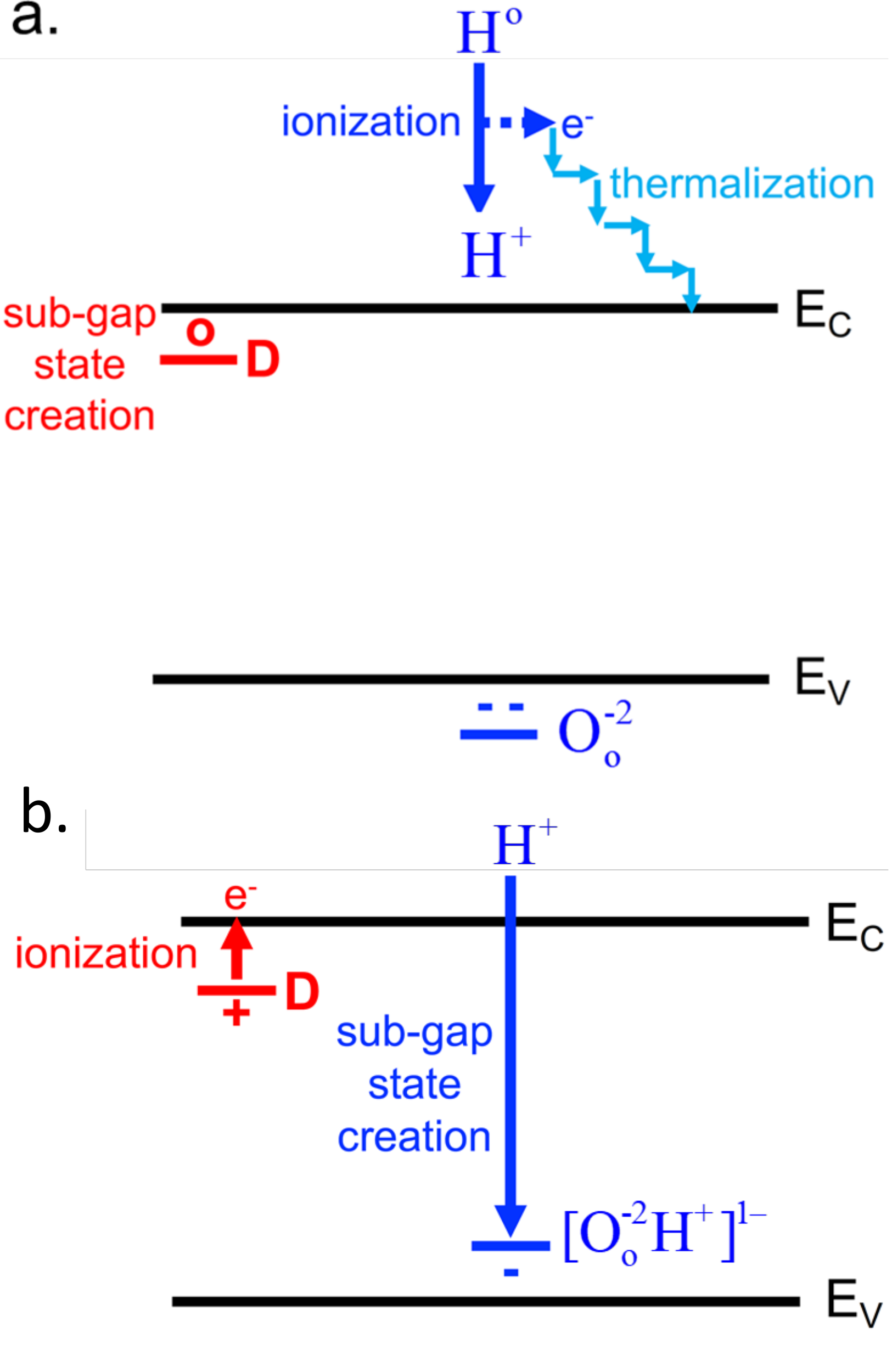}
   \end{tabular}
   \end{center}
   \caption[example] 
%>>>> use \label inside caption to get Fig. number with \ref{}
   { \label{fig6} 
{Normal donor versus hydrogen donor behavior. (a) A normal donor (D, \textit{red}),  creates a neutral, subgap state just below the conduction band minimum when it is incorporated into the lattice, and then in (b-\textit{red}) is positively ionized by thermal emission of an electron into the conduction band. In contrast, hydrogen (H, \textit{blue}) is initially present in the lattice as neutral, non-bonded hydrogen ($\mathrm{H^{0}}$). Then, in (a-\textit{blue}) $\mathrm{H^{0}}$ ionizes to $\mathrm{H^{+}}$ with the ionized electron thermalizing to the bottom of the conduction band. Finally, in (b-\textit{blue}) $\mathrm{H^{+}}$  is incorporated into the lattice, thereby creating (in an oxide) a subgap state just above the valence band mobility edge due to the formation of an $\mathrm{{[{O_{O}^{2-}}{H^+}]}^{1-}}$ defect complex in which an oxygen-on-an-oxygen-site valence band state $\mathrm{O_{O}^{2-}}$ is pushed into the band gap due to forming a complex with $\mathrm{H^{+}}$. } 
}
 \end{figure}

Figure 5 presents a comparison between a normal donor and a hydrogen donor.  As evident from an assessment of Fig. 5, hydrogen donor behavior is significantly different from that of normal donor behavior. When a normal donor (D, indicated as red in Fig. 5) is incorporated into a lattice, it creates a neutral subgap state that is easily thermally ionized, sending its electron into the conduction band. The important point here is the behavioral sequence of a normal donor, involving $\mathrm{\textit{neutral}}$ incorporation $\mathrm{\rightarrow}$ subgap state creation $\mathrm{\rightarrow}$ ionization. In contrast, the behavioral sequence for a hydrogen donor (H, indicated as blue in Fig. 5) is distinctly different, involving ionization $\mathrm{\rightarrow \textit{ionized}}$ incorporation $\mathrm{\rightarrow}$ subgap state creation. More specifically, hydrogen is initially present in the lattice as neutral, non-bonded hydrogen, $\mathrm{H^{0}}$, sitting interstitially in the network. Non-bonded $\mathrm{H^{0}}$ is energetically unstable (metastable) such that it rapidly ionizes to $\mathrm{H^{+}}$ with the ionized electron thermalizing to the bottom of the conduction band (Fig. 5(a)). (If interstitial, non-bonded molecular hydrogen is envisaged is the original source of hydrogen to be incorporated into the lattice, then neutral, non-bonded hydrogen essentially acts as an activated complex facilitating hydrogen ionization, i.e., $\mathrm{{\frac{1}{2}}H_{2} \rightarrow H^{\# 0} \rightarrow H^{+} + e^{-}}$, where the symbol $ \#$ is used to denote an activated complex.) The charged nature of $\mathrm{H^{+}}$ provides a Coulombic driving force for it to bond to an oxygen-on-an-oxygen-site valence band state, $\mathrm{O_{O}^{2-}}$, thereby incorporating itself into the lattice via the formation of an $\mathrm{{[{O_{O}^{2-}}{H^+}]}^{1-}}$ defect complex and creating an associated subgap state energetically positioned just above the valence band mobility edge. Notice that the formation of the $\mathrm{{[{O_{O}^{2-}}{H^+}]}^{1-}}$ defect complex involves energetic re-positioning of the $\mathrm{O_{O}^{2-}}$ valence band state from below to above the valence band mobility edge due to its complexing with $\mathrm{H^{+}}$ (Fig. 5(b)).

In summary, the key difference here is that for a normal donor, subgap state creation occurs first, followed by ionization. In contrast, for the unusual case of a hydrogen donor, ionization occurs first, followed by subgap state creation. Normal donor ionization is controlled by the position of the Fermi level with respect to the donor ionization energy and is thus an equilibrium thermodynamic process. In contrast, since hydrogen donor ionization occurs prior to incorporation into the lattice when subgap states are created, hydrogen donors always remain ionized, independently of the position of the Fermi level; hence, hydrogen donor ionization is a non-equilibrium phenomenon.

%Temperature-dependent measurements were taken on a BCE device that was wire-bonded to a cryostat and cooled down to $140$ K using liquid nitrogen. 

\section{CONCLUSION}

The chemistry and physics of hydrogen incorporation into an a-IGZO TFT channel layer is quite subtle. Experimentally, hydrogen behaves as a donor since its incorporation leads to a negative shift in the TFT turn-on voltage due to an increase in the concentration of free conduction band electrons. However, hydrogen is clearly not a simple donor since its incorporation results in the creation of a Gaussian distribution of subgap electronic states centered at 0.4 eV above the valence band mobility edge and also in an enhancement of the valence band tail state density as witnessed by an increase in the valence band Urbach energy. The hydrogen-induced, Gaussian-distributed, donor-like subgap electronic states arise as a consequence of the negative-U behavior of hydrogen in a-IGZO in which hydrogen ionization precedes its incorporation into the lattice network, resulting in the formation of a $\mathrm{{[{O_{O}^{2-}}{H^+}]}^{1-}}$ defect complex. Furthermore, formation of a $\mathrm{{[{O_{O}^{2-}}{H^+}]}^{1-}}$ defect complex, or equivalently of an OH- bond, leads to positional distortion of the oxygen ion, thus increasing the amount of disorder existing on the anion sublattice, and increasing the valence band tail state density. The charge state of these hydrogen-induced subgap and band tail states are controlled by non-equilibrium and equilibrium thermodynamics, respectively, since they exist in their most positive and neutral charge states, respectively, when the Fermi level is located near to the conduction band mobility edge.

Although a detailed explication of hydrogen incorporation into a-IGZO has been presented, a key question remains unanswered: Why is the concentration of electrically-active hydrogen orders of magnitude smaller than the total concentration of hydrogen as estimated from SIMS analysis? Presumably most of this SIMS-measured hydrogen exists as molecular hydrogen, H$_2$. Moreover, it is likely that H$_2$ is a precursor for electrically-active hydrogen incorporation. But why should such a tiny fraction of this molecular hydrogen become electrically active, and what controls this? Answering these questions requires further investigation.    
\\

\textbf{Data Availability Statement: }The data that support the findings of this study are available from the corresponding author upon reasonable request.

%\textit{Relocated from abstract to save space, incorporate in conclusions?} However, a hydrogen-induced subgap electronic state is non-equilibrium in nature since it persists in its most positive charge state, i.e., $\mathrm{{[{O_{O}^{2-}}{H^+}]}^{1-}}$, even though the position of the Fermi level dictates that the more negative charge state, i.e., $\mathrm{{[{O_{O}^{2-}}{H^+}]}^{2-}}$, would be preferred if equilibrium thermodynamic conditions prevailed.

\bibliography{paperfull2}   %>>>> bibliography data in report.bib

\begin{thebibliography}{40}%
\makeatletter
\providecommand \@ifxundefined [1]{%
 \@ifx{#1\undefined}
}%
\providecommand \@ifnum [1]{%
 \ifnum #1\expandafter \@firstoftwo
 \else \expandafter \@secondoftwo
 \fi
}%
\providecommand \@ifx [1]{%
 \ifx #1\expandafter \@firstoftwo
 \else \expandafter \@secondoftwo
 \fi
}%
\providecommand \natexlab [1]{#1}%
\providecommand \enquote  [1]{``#1''}%
\providecommand \bibnamefont  [1]{#1}%
\providecommand \bibfnamefont [1]{#1}%
\providecommand \citenamefont [1]{#1}%
\providecommand \href@noop [0]{\@secondoftwo}%
\providecommand \href [0]{\begingroup \@sanitize@url \@href}%
\providecommand \@href[1]{\@@startlink{#1}\@@href}%
\providecommand \@@href[1]{\endgroup#1\@@endlink}%
\providecommand \@sanitize@url [0]{\catcode `\\12\catcode `\$12\catcode
  `\&12\catcode `\#12\catcode `\^12\catcode `\_12\catcode `\%12\relax}%
\providecommand \@@startlink[1]{}%
\providecommand \@@endlink[0]{}%
\providecommand \url  [0]{\begingroup\@sanitize@url \@url }%
\providecommand \@url [1]{\endgroup\@href {#1}{\urlprefix }}%
\providecommand \urlprefix  [0]{URL }%
\providecommand \Eprint [0]{\href }%
\providecommand \doibase [0]{http://dx.doi.org/}%
\providecommand \selectlanguage [0]{\@gobble}%
\providecommand \bibinfo  [0]{\@secondoftwo}%
\providecommand \bibfield  [0]{\@secondoftwo}%
\providecommand \translation [1]{[#1]}%
\providecommand \BibitemOpen [0]{}%
\providecommand \bibitemStop [0]{}%
\providecommand \bibitemNoStop [0]{.\EOS\space}%
\providecommand \EOS [0]{\spacefactor3000\relax}%
\providecommand \BibitemShut  [1]{\csname bibitem#1\endcsname}%
\let\auto@bib@innerbib\@empty
%</preamble>
\bibitem [{\citenamefont {{K. Nomura}}\ \emph {et~al.}(2004)\citenamefont {{K.
  Nomura}}, \citenamefont {{H. Ohta}}, \citenamefont {{A. Takagi}},
  \citenamefont {{T. Kamiya}}, \citenamefont {{M. Hirano}},\ and\ \citenamefont
  {Hosono}}]{KnNomura2004}%
  \BibitemOpen
  \bibfield  {author} {\bibinfo {author} {\bibnamefont {{K. Nomura}}}, \bibinfo
  {author} {\bibnamefont {{H. Ohta}}}, \bibinfo {author} {\bibnamefont {{A.
  Takagi}}}, \bibinfo {author} {\bibnamefont {{T. Kamiya}}}, \bibinfo {author}
  {\bibnamefont {{M. Hirano}}}, \ and\ \bibinfo {author} {\bibfnamefont
  {H.}~\bibnamefont {Hosono}},\ }\href@noop {} {\bibfield  {journal} {\bibinfo
  {journal} {Nature}\ }\textbf {\bibinfo {volume} {432}},\ \bibinfo {pages}
  {488} (\bibinfo {year} {2004})}\BibitemShut {NoStop}%
\bibitem [{\citenamefont {Kamiya}, \citenamefont {Nomura},\ and\ \citenamefont
  {Hosono}(2010)}]{Kamiya2010}%
  \BibitemOpen
  \bibfield  {author} {\bibinfo {author} {\bibfnamefont {T.}~\bibnamefont
  {Kamiya}}, \bibinfo {author} {\bibfnamefont {K.}~\bibnamefont {Nomura}}, \
  and\ \bibinfo {author} {\bibfnamefont {H.}~\bibnamefont {Hosono}},\
  }\href@noop {} {\bibfield  {journal} {\bibinfo  {journal} {Science and
  Technology of Advanced Materials}\ }\textbf {\bibinfo {volume} {11}}
  (\bibinfo {year} {2010})}\BibitemShut {NoStop}%
\bibitem [{\citenamefont {Kwon}, \citenamefont {Lee},\ and\ \citenamefont
  {Kim}(2011)}]{Kwon2011}%
  \BibitemOpen
  \bibfield  {author} {\bibinfo {author} {\bibfnamefont {J.~Y.}\ \bibnamefont
  {Kwon}}, \bibinfo {author} {\bibfnamefont {D.~J.}\ \bibnamefont {Lee}}, \
  and\ \bibinfo {author} {\bibfnamefont {K.~B.}\ \bibnamefont {Kim}},\ }\href
  {\doibase 10.1007/s13391-011-0301-x} {\bibfield  {journal} {\bibinfo
  {journal} {Electronic Materials Letters}\ }\textbf {\bibinfo {volume} {7}},\
  \bibinfo {pages} {1} (\bibinfo {year} {2011})}\BibitemShut {NoStop}%
\bibitem [{\citenamefont {Wager}(2020)}]{Wager2020}%
  \BibitemOpen
  \bibfield  {author} {\bibinfo {author} {\bibfnamefont {J.~F.}\ \bibnamefont
  {Wager}},\ }\href {\doibase 10.1002/msid.1098} {\bibfield  {journal}
  {\bibinfo  {journal} {Information Display}\ }\textbf {\bibinfo {volume}
  {36}},\ \bibinfo {pages} {9} (\bibinfo {year} {2020})}\BibitemShut {NoStop}%
\bibitem [{\citenamefont {Vogt}\ \emph {et~al.}(2020)\citenamefont {Vogt},
  \citenamefont {Malmberg}, \citenamefont {Buchanan}, \citenamefont {Mattson},
  \citenamefont {Brandt}, \citenamefont {Fast}, \citenamefont {Cheong},
  \citenamefont {Wager},\ and\ \citenamefont {Graham}}]{Vogt}%
  \BibitemOpen
  \bibfield  {author} {\bibinfo {author} {\bibfnamefont {K.~T.}\ \bibnamefont
  {Vogt}}, \bibinfo {author} {\bibfnamefont {C.~E.}\ \bibnamefont {Malmberg}},
  \bibinfo {author} {\bibfnamefont {J.~C.}\ \bibnamefont {Buchanan}}, \bibinfo
  {author} {\bibfnamefont {G.~W.}\ \bibnamefont {Mattson}}, \bibinfo {author}
  {\bibfnamefont {G.~M.}\ \bibnamefont {Brandt}}, \bibinfo {author}
  {\bibfnamefont {D.~B.}\ \bibnamefont {Fast}}, \bibinfo {author}
  {\bibfnamefont {P.~H.-Y.}\ \bibnamefont {Cheong}}, \bibinfo {author}
  {\bibfnamefont {J.~F.}\ \bibnamefont {Wager}}, \ and\ \bibinfo {author}
  {\bibfnamefont {M.~W.}\ \bibnamefont {Graham}},\ }\href {\doibase
  10.1103/PhysRevResearch.2.033358} {\bibfield  {journal} {\bibinfo  {journal}
  {Phys. Rev. Research}\ }\textbf {\bibinfo {volume} {2}},\ \bibinfo {pages}
  {033358} (\bibinfo {year} {2020})}\BibitemShut {NoStop}%
\bibitem [{\citenamefont {Wager}(2017)}]{Wager2017}%
  \BibitemOpen
  \bibfield  {author} {\bibinfo {author} {\bibfnamefont {J.~F.}\ \bibnamefont
  {Wager}},\ }\href {\doibase 10.1063/1.5008521} {\bibfield  {journal}
  {\bibinfo  {journal} {AIP Advances}\ }\textbf {\bibinfo {volume} {7}},\
  \bibinfo {pages} {1} (\bibinfo {year} {2017})}\BibitemShut {NoStop}%
\bibitem [{\citenamefont {Lee}\ \emph {et~al.}(2009)\citenamefont {Lee},
  \citenamefont {Park}, \citenamefont {Pyo}, \citenamefont {Lee}, \citenamefont
  {Kim}, \citenamefont {Stryakhilev}, \citenamefont {Kim}, \citenamefont
  {Jin},\ and\ \citenamefont {Mo}}]{Lee2009}%
  \BibitemOpen
  \bibfield  {author} {\bibinfo {author} {\bibfnamefont {J.}~\bibnamefont
  {Lee}}, \bibinfo {author} {\bibfnamefont {J.-S.}\ \bibnamefont {Park}},
  \bibinfo {author} {\bibfnamefont {Y.~S.}\ \bibnamefont {Pyo}}, \bibinfo
  {author} {\bibfnamefont {D.~B.}\ \bibnamefont {Lee}}, \bibinfo {author}
  {\bibfnamefont {E.~H.}\ \bibnamefont {Kim}}, \bibinfo {author} {\bibfnamefont
  {D.}~\bibnamefont {Stryakhilev}}, \bibinfo {author} {\bibfnamefont {T.~W.}\
  \bibnamefont {Kim}}, \bibinfo {author} {\bibfnamefont {D.~U.}\ \bibnamefont
  {Jin}}, \ and\ \bibinfo {author} {\bibfnamefont {Y.-G.}\ \bibnamefont {Mo}},\
  }\href {\doibase 10.1063/1.3232179} {\bibfield  {journal} {\bibinfo
  {journal} {Applied Physics Letters}\ }\textbf {\bibinfo {volume} {95}},\
  \bibinfo {pages} {123502} (\bibinfo {year} {2009})}\BibitemShut {NoStop}%
\bibitem [{\citenamefont {Miyase}\ \emph {et~al.}(2014)\citenamefont {Miyase},
  \citenamefont {Watanabe}, \citenamefont {Sakaguchi}, \citenamefont {Ohashi},
  \citenamefont {Domen}, \citenamefont {Nomura}, \citenamefont {Hiramatsu},
  \citenamefont {Kumomi}, \citenamefont {Hosono},\ and\ \citenamefont
  {Kamiya}}]{Miyase2014}%
  \BibitemOpen
  \bibfield  {author} {\bibinfo {author} {\bibfnamefont {T.}~\bibnamefont
  {Miyase}}, \bibinfo {author} {\bibfnamefont {K.}~\bibnamefont {Watanabe}},
  \bibinfo {author} {\bibfnamefont {I.}~\bibnamefont {Sakaguchi}}, \bibinfo
  {author} {\bibfnamefont {N.}~\bibnamefont {Ohashi}}, \bibinfo {author}
  {\bibfnamefont {K.}~\bibnamefont {Domen}}, \bibinfo {author} {\bibfnamefont
  {K.}~\bibnamefont {Nomura}}, \bibinfo {author} {\bibfnamefont
  {H.}~\bibnamefont {Hiramatsu}}, \bibinfo {author} {\bibfnamefont
  {H.}~\bibnamefont {Kumomi}}, \bibinfo {author} {\bibfnamefont
  {H.}~\bibnamefont {Hosono}}, \ and\ \bibinfo {author} {\bibfnamefont
  {T.}~\bibnamefont {Kamiya}},\ }\href {\doibase 10.1149/2.015409jss}
  {\bibfield  {journal} {\bibinfo  {journal} {ECS Journal of Solid State
  Science and Technology}\ }\textbf {\bibinfo {volume} {3}},\ \bibinfo {pages}
  {Q3085} (\bibinfo {year} {2014})}\BibitemShut {NoStop}%
\bibitem [{\citenamefont {Bang}, \citenamefont {Matsuishi},\ and\ \citenamefont
  {Hosono}(2017)}]{Bang2017}%
  \BibitemOpen
  \bibfield  {author} {\bibinfo {author} {\bibfnamefont {J.}~\bibnamefont
  {Bang}}, \bibinfo {author} {\bibfnamefont {S.}~\bibnamefont {Matsuishi}}, \
  and\ \bibinfo {author} {\bibfnamefont {H.}~\bibnamefont {Hosono}},\ }\href
  {\doibase 10.1063/1.4985627} {\bibfield  {journal} {\bibinfo  {journal}
  {Applied Physics Letters}\ }\textbf {\bibinfo {volume} {110}} (\bibinfo
  {year} {2017}),\ 10.1063/1.4985627}\BibitemShut {NoStop}%
\bibitem [{\citenamefont {Li}, \citenamefont {Guo},\ and\ \citenamefont
  {Robertson}(2017)}]{Li2017}%
  \BibitemOpen
  \bibfield  {author} {\bibinfo {author} {\bibfnamefont {H.}~\bibnamefont
  {Li}}, \bibinfo {author} {\bibfnamefont {Y.}~\bibnamefont {Guo}}, \ and\
  \bibinfo {author} {\bibfnamefont {J.}~\bibnamefont {Robertson}},\ }\href
  {\doibase 10.1038/s41598-017-17290-5} {\bibfield  {journal} {\bibinfo
  {journal} {Scientific Reports}\ }\textbf {\bibinfo {volume} {7}},\ \bibinfo
  {pages} {1} (\bibinfo {year} {2017})}\BibitemShut {NoStop}%
\bibitem [{\citenamefont {Song}\ \emph {et~al.}(2017)\citenamefont {Song},
  \citenamefont {Park}, \citenamefont {Chung}, \citenamefont {Rim},
  \citenamefont {Son}, \citenamefont {Lim},\ and\ \citenamefont
  {Chu}}]{Song2017}%
  \BibitemOpen
  \bibfield  {author} {\bibinfo {author} {\bibfnamefont {A.}~\bibnamefont
  {Song}}, \bibinfo {author} {\bibfnamefont {H.-W.}\ \bibnamefont {Park}},
  \bibinfo {author} {\bibfnamefont {K.-B.}\ \bibnamefont {Chung}}, \bibinfo
  {author} {\bibfnamefont {Y.~S.}\ \bibnamefont {Rim}}, \bibinfo {author}
  {\bibfnamefont {K.~S.}\ \bibnamefont {Son}}, \bibinfo {author} {\bibfnamefont
  {J.~H.}\ \bibnamefont {Lim}}, \ and\ \bibinfo {author} {\bibfnamefont
  {H.~Y.}\ \bibnamefont {Chu}},\ }\href {\doibase 10.1063/1.5003186} {\bibfield
   {journal} {\bibinfo  {journal} {Applied Physics Letters}\ }\textbf {\bibinfo
  {volume} {111}},\ \bibinfo {pages} {243507} (\bibinfo {year}
  {2017})}\BibitemShut {NoStop}%
\bibitem [{\citenamefont {Han}\ \emph {et~al.}(2017)\citenamefont {Han},
  \citenamefont {Ok}, \citenamefont {Cho}, \citenamefont {Oh},\ and\
  \citenamefont {Park}}]{Han2017}%
  \BibitemOpen
  \bibfield  {author} {\bibinfo {author} {\bibfnamefont {K.-L.}\ \bibnamefont
  {Han}}, \bibinfo {author} {\bibfnamefont {K.-C.}\ \bibnamefont {Ok}},
  \bibinfo {author} {\bibfnamefont {H.-S.}\ \bibnamefont {Cho}}, \bibinfo
  {author} {\bibfnamefont {S.}~\bibnamefont {Oh}}, \ and\ \bibinfo {author}
  {\bibfnamefont {J.-S.}\ \bibnamefont {Park}},\ }\href {\doibase
  10.1063/1.4997926} {\bibfield  {journal} {\bibinfo  {journal} {Applied
  Physics Letters}\ }\textbf {\bibinfo {volume} {111}},\ \bibinfo {pages}
  {063502} (\bibinfo {year} {2017})}\BibitemShut {NoStop}%
\bibitem [{\citenamefont {Abliz}\ \emph {et~al.}(2017)\citenamefont {Abliz},
  \citenamefont {Gao}, \citenamefont {Wan}, \citenamefont {Liu}, \citenamefont
  {Xu}, \citenamefont {Liu}, \citenamefont {Jiang}, \citenamefont {Li},
  \citenamefont {Chen}, \citenamefont {Guo}, \citenamefont {Li},\ and\
  \citenamefont {Liao}}]{Abliz2017}%
  \BibitemOpen
  \bibfield  {author} {\bibinfo {author} {\bibfnamefont {A.}~\bibnamefont
  {Abliz}}, \bibinfo {author} {\bibfnamefont {Q.}~\bibnamefont {Gao}}, \bibinfo
  {author} {\bibfnamefont {D.}~\bibnamefont {Wan}}, \bibinfo {author}
  {\bibfnamefont {X.}~\bibnamefont {Liu}}, \bibinfo {author} {\bibfnamefont
  {L.}~\bibnamefont {Xu}}, \bibinfo {author} {\bibfnamefont {C.}~\bibnamefont
  {Liu}}, \bibinfo {author} {\bibfnamefont {C.}~\bibnamefont {Jiang}}, \bibinfo
  {author} {\bibfnamefont {X.}~\bibnamefont {Li}}, \bibinfo {author}
  {\bibfnamefont {H.}~\bibnamefont {Chen}}, \bibinfo {author} {\bibfnamefont
  {T.}~\bibnamefont {Guo}}, \bibinfo {author} {\bibfnamefont {J.}~\bibnamefont
  {Li}}, \ and\ \bibinfo {author} {\bibfnamefont {L.}~\bibnamefont {Liao}},\
  }\href {\doibase 10.1021/acsami.6b15275} {\bibfield  {journal} {\bibinfo
  {journal} {ACS Applied Materials and Interfaces}\ }\textbf {\bibinfo {volume}
  {9}},\ \bibinfo {pages} {10798} (\bibinfo {year} {2017})}\BibitemShut
  {NoStop}%
\bibitem [{\citenamefont {Nam}\ \emph {et~al.}(2018)\citenamefont {Nam},
  \citenamefont {Kim}, \citenamefont {Cho},\ and\ \citenamefont {{Ko
  Park}}}]{Nam2018}%
  \BibitemOpen
  \bibfield  {author} {\bibinfo {author} {\bibfnamefont {Y.}~\bibnamefont
  {Nam}}, \bibinfo {author} {\bibfnamefont {H.~O.}\ \bibnamefont {Kim}},
  \bibinfo {author} {\bibfnamefont {S.~H.}\ \bibnamefont {Cho}}, \ and\
  \bibinfo {author} {\bibfnamefont {S.~H.}\ \bibnamefont {{Ko Park}}},\ }\href
  {\doibase 10.1039/c7ra12841j} {\bibfield  {journal} {\bibinfo  {journal} {RSC
  Advances}\ }\textbf {\bibinfo {volume} {8}},\ \bibinfo {pages} {5622}
  (\bibinfo {year} {2018})}\BibitemShut {NoStop}%
\bibitem [{\citenamefont {Felizco}\ \emph {et~al.}(2020)\citenamefont
  {Felizco}, \citenamefont {Uenuma}, \citenamefont {Ishikawa},\ and\
  \citenamefont {Uraoka}}]{Felizco2020}%
  \BibitemOpen
  \bibfield  {author} {\bibinfo {author} {\bibfnamefont {J.~C.}\ \bibnamefont
  {Felizco}}, \bibinfo {author} {\bibfnamefont {M.}~\bibnamefont {Uenuma}},
  \bibinfo {author} {\bibfnamefont {Y.}~\bibnamefont {Ishikawa}}, \ and\
  \bibinfo {author} {\bibfnamefont {Y.}~\bibnamefont {Uraoka}},\ }\href
  {\doibase 10.1016/j.apsusc.2020.146791} {\bibfield  {journal} {\bibinfo
  {journal} {Applied Surface Science}\ }\textbf {\bibinfo {volume} {527}},\
  \bibinfo {pages} {146791} (\bibinfo {year} {2020})}\BibitemShut {NoStop}%
\bibitem [{\citenamefont {Chen}\ \emph
  {et~al.}(2020{\natexlab{a}})\citenamefont {Chen}, \citenamefont {Chen},
  \citenamefont {Zhou}, \citenamefont {Chen}, \citenamefont {Kuo},
  \citenamefont {Shih}, \citenamefont {Su}, \citenamefont {Yang}, \citenamefont
  {Huang}, \citenamefont {Shih}, \citenamefont {Lai},\ and\ \citenamefont
  {Chang}}]{Chen2020}%
  \BibitemOpen
  \bibfield  {author} {\bibinfo {author} {\bibfnamefont {H.-C.}\ \bibnamefont
  {Chen}}, \bibinfo {author} {\bibfnamefont {J.-J.}\ \bibnamefont {Chen}},
  \bibinfo {author} {\bibfnamefont {K.-J.}\ \bibnamefont {Zhou}}, \bibinfo
  {author} {\bibfnamefont {G.-F.}\ \bibnamefont {Chen}}, \bibinfo {author}
  {\bibfnamefont {C.-W.}\ \bibnamefont {Kuo}}, \bibinfo {author} {\bibfnamefont
  {Y.-S.}\ \bibnamefont {Shih}}, \bibinfo {author} {\bibfnamefont {W.-C.}\
  \bibnamefont {Su}}, \bibinfo {author} {\bibfnamefont {C.-C.}\ \bibnamefont
  {Yang}}, \bibinfo {author} {\bibfnamefont {H.-C.}\ \bibnamefont {Huang}},
  \bibinfo {author} {\bibfnamefont {C.-C.}\ \bibnamefont {Shih}}, \bibinfo
  {author} {\bibfnamefont {W.-C.}\ \bibnamefont {Lai}}, \ and\ \bibinfo
  {author} {\bibfnamefont {T.-C.}\ \bibnamefont {Chang}},\ }\href {\doibase
  10.1109/TED.2020.2998101} {\bibfield  {journal} {\bibinfo  {journal} {IEEE
  Transactions on Electron Devices}\ }\textbf {\bibinfo {volume} {67}},\
  \bibinfo {pages} {3123} (\bibinfo {year} {2020}{\natexlab{a}})}\BibitemShut
  {NoStop}%
\bibitem [{\citenamefont {Wang}\ \emph {et~al.}(2020)\citenamefont {Wang},
  \citenamefont {Shao}, \citenamefont {Wu}, \citenamefont {Zhang},
  \citenamefont {Li}, \citenamefont {Liu}, \citenamefont {Zhang},\ and\
  \citenamefont {Ding}}]{Wang2020}%
  \BibitemOpen
  \bibfield  {author} {\bibinfo {author} {\bibfnamefont {X.-L.}\ \bibnamefont
  {Wang}}, \bibinfo {author} {\bibfnamefont {Y.}~\bibnamefont {Shao}}, \bibinfo
  {author} {\bibfnamefont {X.}~\bibnamefont {Wu}}, \bibinfo {author}
  {\bibfnamefont {M.-N.}\ \bibnamefont {Zhang}}, \bibinfo {author}
  {\bibfnamefont {L.}~\bibnamefont {Li}}, \bibinfo {author} {\bibfnamefont
  {W.-J.}\ \bibnamefont {Liu}}, \bibinfo {author} {\bibfnamefont {D.~W.}\
  \bibnamefont {Zhang}}, \ and\ \bibinfo {author} {\bibfnamefont {S.-J.}\
  \bibnamefont {Ding}},\ }\href {\doibase 10.1039/c9ra09646a} {\bibfield
  {journal} {\bibinfo  {journal} {{RSC} Advances}\ }\textbf {\bibinfo {volume}
  {10}},\ \bibinfo {pages} {3572} (\bibinfo {year} {2020})}\BibitemShut
  {NoStop}%
\bibitem [{\citenamefont {Abliz}(2020)}]{Abliz2020}%
  \BibitemOpen
  \bibfield  {author} {\bibinfo {author} {\bibfnamefont {A.}~\bibnamefont
  {Abliz}},\ }\href {\doibase 10.1016/j.jallcom.2020.154694} {\bibfield
  {journal} {\bibinfo  {journal} {Journal of Alloys and Compounds}\ }\textbf
  {\bibinfo {volume} {831}},\ \bibinfo {pages} {154694} (\bibinfo {year}
  {2020})}\BibitemShut {NoStop}%
\bibitem [{\citenamefont {Corsino}\ \emph {et~al.}(2020)\citenamefont
  {Corsino}, \citenamefont {Bermundo}, \citenamefont {Fujii}, \citenamefont
  {Takahashi}, \citenamefont {Ishikawa},\ and\ \citenamefont
  {Uraoka}}]{Corsino2020}%
  \BibitemOpen
  \bibfield  {author} {\bibinfo {author} {\bibfnamefont {D.~C.}\ \bibnamefont
  {Corsino}}, \bibinfo {author} {\bibfnamefont {J.~P.~S.}\ \bibnamefont
  {Bermundo}}, \bibinfo {author} {\bibfnamefont {M.~N.}\ \bibnamefont {Fujii}},
  \bibinfo {author} {\bibfnamefont {K.}~\bibnamefont {Takahashi}}, \bibinfo
  {author} {\bibfnamefont {Y.}~\bibnamefont {Ishikawa}}, \ and\ \bibinfo
  {author} {\bibfnamefont {Y.}~\bibnamefont {Uraoka}},\ }\href {\doibase
  10.1088/1361-6463/ab6e97} {\bibfield  {journal} {\bibinfo  {journal} {Journal
  of Physics D: Applied Physics}\ }\textbf {\bibinfo {volume} {53}},\ \bibinfo
  {pages} {165103} (\bibinfo {year} {2020})}\BibitemShut {NoStop}%
\bibitem [{\citenamefont {Prasad}\ \emph {et~al.}(2021)\citenamefont {Prasad},
  \citenamefont {Mohanty}, \citenamefont {Wu}, \citenamefont {Yu},\ and\
  \citenamefont {Chang}}]{Prasad2021}%
  \BibitemOpen
  \bibfield  {author} {\bibinfo {author} {\bibfnamefont {O.~K.}\ \bibnamefont
  {Prasad}}, \bibinfo {author} {\bibfnamefont {S.~K.}\ \bibnamefont {Mohanty}},
  \bibinfo {author} {\bibfnamefont {C.~H.}\ \bibnamefont {Wu}}, \bibinfo
  {author} {\bibfnamefont {T.~Y.}\ \bibnamefont {Yu}}, \ and\ \bibinfo {author}
  {\bibfnamefont {K.~M.}\ \bibnamefont {Chang}},\ }\href {\doibase
  10.1088/1361-6528/ac0cb0} {\bibfield  {journal} {\bibinfo  {journal}
  {Nanotechnology}\ }\textbf {\bibinfo {volume} {32}},\ \bibinfo {pages}
  {395203} (\bibinfo {year} {2021})}\BibitemShut {NoStop}%
\bibitem [{\citenamefont {Magari}\ and\ \citenamefont
  {Furura}(2021)}]{Magari2021}%
  \BibitemOpen
  \bibfield  {author} {\bibinfo {author} {\bibfnamefont {Y.}~\bibnamefont
  {Magari}}\ and\ \bibinfo {author} {\bibfnamefont {M.}~\bibnamefont
  {Furura}},\ }\href {\doibase 10.35848/1347-4065/abd9d2} {\bibfield  {journal}
  {\bibinfo  {journal} {Japanese Journal of Applied Physics}\ }\textbf
  {\bibinfo {volume} {60}},\ \bibinfo {pages} {SBBM04} (\bibinfo {year}
  {2021})}\BibitemShut {NoStop}%
\bibitem [{\citenamefont {{De Jamblinne De Meux}}\ \emph
  {et~al.}(2018)\citenamefont {{De Jamblinne De Meux}}, \citenamefont
  {Pourtois}, \citenamefont {Genoe},\ and\ \citenamefont
  {Heremans}}]{DeJamblinneDeMeux2018}%
  \BibitemOpen
  \bibfield  {author} {\bibinfo {author} {\bibfnamefont {A.}~\bibnamefont {{De
  Jamblinne De Meux}}}, \bibinfo {author} {\bibfnamefont {G.}~\bibnamefont
  {Pourtois}}, \bibinfo {author} {\bibfnamefont {J.}~\bibnamefont {Genoe}}, \
  and\ \bibinfo {author} {\bibfnamefont {P.}~\bibnamefont {Heremans}},\ }\href
  {\doibase 10.1103/PhysRevApplied.9.054039} {\bibfield  {journal} {\bibinfo
  {journal} {Physical Review Applied}\ }\textbf {\bibinfo {volume} {9}},\
  \bibinfo {pages} {54039} (\bibinfo {year} {2018})}\BibitemShut {NoStop}%
\bibitem [{\citenamefont {Lee}\ \emph {et~al.}(2021)\citenamefont {Lee},
  \citenamefont {Nam}, \citenamefont {Seo}, \citenamefont {Yoon}, \citenamefont
  {Oh}, \citenamefont {Lee}, \citenamefont {Yoo}, \citenamefont {Kim},
  \citenamefont {Choi}, \citenamefont {Im}, \citenamefont {Yang}, \citenamefont
  {Choi}, \citenamefont {Yoo}, \citenamefont {jin Kim},\ and\ \citenamefont
  {Kim}}]{Lee2021}%
  \BibitemOpen
  \bibfield  {author} {\bibinfo {author} {\bibfnamefont {Y.}~\bibnamefont
  {Lee}}, \bibinfo {author} {\bibfnamefont {T.}~\bibnamefont {Nam}}, \bibinfo
  {author} {\bibfnamefont {S.}~\bibnamefont {Seo}}, \bibinfo {author}
  {\bibfnamefont {H.}~\bibnamefont {Yoon}}, \bibinfo {author} {\bibfnamefont
  {I.-K.}\ \bibnamefont {Oh}}, \bibinfo {author} {\bibfnamefont {C.~H.}\
  \bibnamefont {Lee}}, \bibinfo {author} {\bibfnamefont {H.}~\bibnamefont
  {Yoo}}, \bibinfo {author} {\bibfnamefont {H.~J.}\ \bibnamefont {Kim}},
  \bibinfo {author} {\bibfnamefont {W.}~\bibnamefont {Choi}}, \bibinfo {author}
  {\bibfnamefont {S.}~\bibnamefont {Im}}, \bibinfo {author} {\bibfnamefont
  {J.~Y.}\ \bibnamefont {Yang}}, \bibinfo {author} {\bibfnamefont {D.~W.}\
  \bibnamefont {Choi}}, \bibinfo {author} {\bibfnamefont {C.}~\bibnamefont
  {Yoo}}, \bibinfo {author} {\bibfnamefont {H.}~\bibnamefont {jin Kim}}, \ and\
  \bibinfo {author} {\bibfnamefont {H.}~\bibnamefont {Kim}},\ }\href {\doibase
  10.1021/acsami.1c02597} {\bibfield  {journal} {\bibinfo  {journal} {{ACS}
  Applied Materials {\&} Interfaces}\ }\textbf {\bibinfo {volume} {13}},\
  \bibinfo {pages} {20349} (\bibinfo {year} {2021})}\BibitemShut {NoStop}%
\bibitem [{\citenamefont {Jeong}\ \emph {et~al.}(2020)\citenamefont {Jeong},
  \citenamefont {Jeong}, \citenamefont {Choi}, \citenamefont {Kim},\ and\
  \citenamefont {Park}}]{Jeong2020}%
  \BibitemOpen
  \bibfield  {author} {\bibinfo {author} {\bibfnamefont {S.-G.}\ \bibnamefont
  {Jeong}}, \bibinfo {author} {\bibfnamefont {H.-J.}\ \bibnamefont {Jeong}},
  \bibinfo {author} {\bibfnamefont {W.-H.}\ \bibnamefont {Choi}}, \bibinfo
  {author} {\bibfnamefont {K.}~\bibnamefont {Kim}}, \ and\ \bibinfo {author}
  {\bibfnamefont {J.-S.}\ \bibnamefont {Park}},\ }\href {\doibase
  10.1109/TED.2020.3017145} {\bibfield  {journal} {\bibinfo  {journal} {IEEE
  Transactions on Electron Devices}\ }\textbf {\bibinfo {volume} {67}},\
  \bibinfo {pages} {4250} (\bibinfo {year} {2020})}\BibitemShut {NoStop}%
\bibitem [{\citenamefont {Chen}\ \emph
  {et~al.}(2020{\natexlab{b}})\citenamefont {Chen}, \citenamefont {Chen},
  \citenamefont {Tu}, \citenamefont {Zhou}, \citenamefont {Kuo}, \citenamefont
  {Su}, \citenamefont {Hung}, \citenamefont {Shih}, \citenamefont {Huang},
  \citenamefont {Tsai}, \citenamefont {Huang}, \citenamefont {Lai},\ and\
  \citenamefont {Chang}}]{Chen2020_hump}%
  \BibitemOpen
  \bibfield  {author} {\bibinfo {author} {\bibfnamefont {H.-C.}\ \bibnamefont
  {Chen}}, \bibinfo {author} {\bibfnamefont {J.-J.}\ \bibnamefont {Chen}},
  \bibinfo {author} {\bibfnamefont {Y.-F.}\ \bibnamefont {Tu}}, \bibinfo
  {author} {\bibfnamefont {K.-J.}\ \bibnamefont {Zhou}}, \bibinfo {author}
  {\bibfnamefont {C.-W.}\ \bibnamefont {Kuo}}, \bibinfo {author} {\bibfnamefont
  {W.-C.}\ \bibnamefont {Su}}, \bibinfo {author} {\bibfnamefont {Y.-H.}\
  \bibnamefont {Hung}}, \bibinfo {author} {\bibfnamefont {Y.-S.}\ \bibnamefont
  {Shih}}, \bibinfo {author} {\bibfnamefont {H.-C.}\ \bibnamefont {Huang}},
  \bibinfo {author} {\bibfnamefont {T.-M.}\ \bibnamefont {Tsai}}, \bibinfo
  {author} {\bibfnamefont {J.-W.}\ \bibnamefont {Huang}}, \bibinfo {author}
  {\bibfnamefont {W.-C.}\ \bibnamefont {Lai}}, \ and\ \bibinfo {author}
  {\bibfnamefont {T.-C.}\ \bibnamefont {Chang}},\ }\href {\doibase
  10.1109/TED.2020.2994539} {\bibfield  {journal} {\bibinfo  {journal} {IEEE
  Transactions on Electron Devices}\ }\textbf {\bibinfo {volume} {67}},\
  \bibinfo {pages} {2807} (\bibinfo {year} {2020}{\natexlab{b}})}\BibitemShut
  {NoStop}%
\bibitem [{\citenamefont {Liu}\ \emph {et~al.}(2021)\citenamefont {Liu},
  \citenamefont {Qin}, \citenamefont {Liu}, \citenamefont {Wei}, \citenamefont
  {Wang},\ and\ \citenamefont {Zhao}}]{Liu2021}%
  \BibitemOpen
  \bibfield  {author} {\bibinfo {author} {\bibfnamefont {C.}~\bibnamefont
  {Liu}}, \bibinfo {author} {\bibfnamefont {H.}~\bibnamefont {Qin}}, \bibinfo
  {author} {\bibfnamefont {Y.}~\bibnamefont {Liu}}, \bibinfo {author}
  {\bibfnamefont {S.}~\bibnamefont {Wei}}, \bibinfo {author} {\bibfnamefont
  {H.}~\bibnamefont {Wang}}, \ and\ \bibinfo {author} {\bibfnamefont
  {Y.}~\bibnamefont {Zhao}},\ }\href {\doibase 10.1016/j.mssp.2020.105390}
  {\bibfield  {journal} {\bibinfo  {journal} {Materials Science in
  Semiconductor Processing}\ }\textbf {\bibinfo {volume} {121}},\ \bibinfo
  {pages} {105390} (\bibinfo {year} {2021})}\BibitemShut {NoStop}%
\bibitem [{\citenamefont {Kim}\ \emph {et~al.}(2012)\citenamefont {Kim},
  \citenamefont {Park}, \citenamefont {Jung}, \citenamefont {Son},
  \citenamefont {Lee}, \citenamefont {Lee}, \citenamefont {Jeong},
  \citenamefont {Mo}, \citenamefont {Son}, \citenamefont {Ryu}, \citenamefont
  {Lee},\ and\ \citenamefont {Jeong}}]{Kim2012}%
  \BibitemOpen
  \bibfield  {author} {\bibinfo {author} {\bibfnamefont {H.~J.}\ \bibnamefont
  {Kim}}, \bibinfo {author} {\bibfnamefont {S.~Y.}\ \bibnamefont {Park}},
  \bibinfo {author} {\bibfnamefont {H.~Y.}\ \bibnamefont {Jung}}, \bibinfo
  {author} {\bibfnamefont {B.~G.}\ \bibnamefont {Son}}, \bibinfo {author}
  {\bibfnamefont {C.-K.}\ \bibnamefont {Lee}}, \bibinfo {author} {\bibfnamefont
  {C.-K.}\ \bibnamefont {Lee}}, \bibinfo {author} {\bibfnamefont {J.~H.}\
  \bibnamefont {Jeong}}, \bibinfo {author} {\bibfnamefont {Y.-G.}\ \bibnamefont
  {Mo}}, \bibinfo {author} {\bibfnamefont {K.~S.}\ \bibnamefont {Son}},
  \bibinfo {author} {\bibfnamefont {M.~K.}\ \bibnamefont {Ryu}}, \bibinfo
  {author} {\bibfnamefont {S.}~\bibnamefont {Lee}}, \ and\ \bibinfo {author}
  {\bibfnamefont {J.~K.}\ \bibnamefont {Jeong}},\ }\href {\doibase
  10.1088/0022-3727/46/5/055104} {\bibfield  {journal} {\bibinfo  {journal}
  {Journal of Physics D: Applied Physics}\ }\textbf {\bibinfo {volume} {46}},\
  \bibinfo {pages} {055104} (\bibinfo {year} {2012})}\BibitemShut {NoStop}%
\bibitem [{\citenamefont {Kim}\ \emph {et~al.}(2019)\citenamefont {Kim},
  \citenamefont {Kim}, \citenamefont {Lee}, \citenamefont {Park}, \citenamefont
  {Chang}, \citenamefont {Kim},\ and\ \citenamefont {Choi}}]{Kim2019}%
  \BibitemOpen
  \bibfield  {author} {\bibinfo {author} {\bibfnamefont {D.-G.}\ \bibnamefont
  {Kim}}, \bibinfo {author} {\bibfnamefont {J.-U.}\ \bibnamefont {Kim}},
  \bibinfo {author} {\bibfnamefont {J.-S.}\ \bibnamefont {Lee}}, \bibinfo
  {author} {\bibfnamefont {K.-S.}\ \bibnamefont {Park}}, \bibinfo {author}
  {\bibfnamefont {Y.-G.}\ \bibnamefont {Chang}}, \bibinfo {author}
  {\bibfnamefont {M.-H.}\ \bibnamefont {Kim}}, \ and\ \bibinfo {author}
  {\bibfnamefont {D.-K.}\ \bibnamefont {Choi}},\ }\href {\doibase
  10.1039/c9ra03053k} {\bibfield  {journal} {\bibinfo  {journal} {RSC
  Advances}\ }\textbf {\bibinfo {volume} {9}},\ \bibinfo {pages} {20865}
  (\bibinfo {year} {2019})}\BibitemShut {NoStop}%
\bibitem [{\citenamefont {Song}\ \emph {et~al.}(2021)\citenamefont {Song},
  \citenamefont {Hong}, \citenamefont {Son}, \citenamefont {Lim},\ and\
  \citenamefont {Chung}}]{Song2021}%
  \BibitemOpen
  \bibfield  {author} {\bibinfo {author} {\bibfnamefont {A.}~\bibnamefont
  {Song}}, \bibinfo {author} {\bibfnamefont {H.~M.}\ \bibnamefont {Hong}},
  \bibinfo {author} {\bibfnamefont {K.~S.}\ \bibnamefont {Son}}, \bibinfo
  {author} {\bibfnamefont {J.~H.}\ \bibnamefont {Lim}}, \ and\ \bibinfo
  {author} {\bibfnamefont {K.-B.}\ \bibnamefont {Chung}},\ }\href {\doibase
  10.1109/ted.2021.3074120} {\bibfield  {journal} {\bibinfo  {journal} {IEEE
  Transactions on Electron Devices}\ }\textbf {\bibinfo {volume} {68}},\
  \bibinfo {pages} {2723} (\bibinfo {year} {2021})}\BibitemShut {NoStop}%
\bibitem [{\citenamefont {Kumomi}\ \emph {et~al.}(2009)\citenamefont {Kumomi},
  \citenamefont {Yaginuma}, \citenamefont {Omura}, \citenamefont {Goyal},
  \citenamefont {Sato}, \citenamefont {Watanabe}, \citenamefont {Shimada},
  \citenamefont {Kaji}, \citenamefont {Takahashi}, \citenamefont {Ofuji},
  \citenamefont {Watanabe}, \citenamefont {Itagaki}, \citenamefont {Shimizu},
  \citenamefont {Abe}, \citenamefont {Tateishi}, \citenamefont {Yabuta},
  \citenamefont {Iwasaki}, \citenamefont {Hayashi}, \citenamefont {Aiba},\ and\
  \citenamefont {Sano}}]{Kumomi2009}%
  \BibitemOpen
  \bibfield  {author} {\bibinfo {author} {\bibfnamefont {H.}~\bibnamefont
  {Kumomi}}, \bibinfo {author} {\bibfnamefont {S.}~\bibnamefont {Yaginuma}},
  \bibinfo {author} {\bibfnamefont {H.}~\bibnamefont {Omura}}, \bibinfo
  {author} {\bibfnamefont {A.}~\bibnamefont {Goyal}}, \bibinfo {author}
  {\bibfnamefont {A.}~\bibnamefont {Sato}}, \bibinfo {author} {\bibfnamefont
  {M.}~\bibnamefont {Watanabe}}, \bibinfo {author} {\bibfnamefont
  {M.}~\bibnamefont {Shimada}}, \bibinfo {author} {\bibfnamefont
  {N.}~\bibnamefont {Kaji}}, \bibinfo {author} {\bibfnamefont {K.}~\bibnamefont
  {Takahashi}}, \bibinfo {author} {\bibfnamefont {M.}~\bibnamefont {Ofuji}},
  \bibinfo {author} {\bibfnamefont {T.}~\bibnamefont {Watanabe}}, \bibinfo
  {author} {\bibfnamefont {N.}~\bibnamefont {Itagaki}}, \bibinfo {author}
  {\bibfnamefont {H.}~\bibnamefont {Shimizu}}, \bibinfo {author} {\bibfnamefont
  {K.}~\bibnamefont {Abe}}, \bibinfo {author} {\bibfnamefont {Y.}~\bibnamefont
  {Tateishi}}, \bibinfo {author} {\bibfnamefont {H.}~\bibnamefont {Yabuta}},
  \bibinfo {author} {\bibfnamefont {T.}~\bibnamefont {Iwasaki}}, \bibinfo
  {author} {\bibfnamefont {R.}~\bibnamefont {Hayashi}}, \bibinfo {author}
  {\bibfnamefont {T.}~\bibnamefont {Aiba}}, \ and\ \bibinfo {author}
  {\bibfnamefont {M.}~\bibnamefont {Sano}},\ }\href {\doibase
  10.1109/jdt.2009.2025521} {\bibfield  {journal} {\bibinfo  {journal} {Journal
  of Display Technology}\ }\textbf {\bibinfo {volume} {5}},\ \bibinfo {pages}
  {531} (\bibinfo {year} {2009})}\BibitemShut {NoStop}%
\bibitem [{\citenamefont {Nomura}, \citenamefont {Kamiya},\ and\ \citenamefont
  {Hosono}(2013)}]{Nomura2013}%
  \BibitemOpen
  \bibfield  {author} {\bibinfo {author} {\bibfnamefont {K.}~\bibnamefont
  {Nomura}}, \bibinfo {author} {\bibfnamefont {T.}~\bibnamefont {Kamiya}}, \
  and\ \bibinfo {author} {\bibfnamefont {H.}~\bibnamefont {Hosono}},\ }\href
  {\doibase 10.1149/2.011301jss} {\bibfield  {journal} {\bibinfo  {journal}
  {ECS Journal of Solid State Science and Technology}\ }\textbf {\bibinfo
  {volume} {2}},\ \bibinfo {pages} {P5} (\bibinfo {year} {2013})}\BibitemShut
  {NoStop}%
\bibitem [{\citenamefont {Jang}\ \emph {et~al.}(2021)\citenamefont {Jang},
  \citenamefont {Ko}, \citenamefont {Choi}, \citenamefont {Kim},\ and\
  \citenamefont {Kim}}]{Jang2021}%
  \BibitemOpen
  \bibfield  {author} {\bibinfo {author} {\bibfnamefont {J.~T.}\ \bibnamefont
  {Jang}}, \bibinfo {author} {\bibfnamefont {D.}~\bibnamefont {Ko}}, \bibinfo
  {author} {\bibfnamefont {S.-J.}\ \bibnamefont {Choi}}, \bibinfo {author}
  {\bibfnamefont {D.~M.}\ \bibnamefont {Kim}}, \ and\ \bibinfo {author}
  {\bibfnamefont {D.~H.}\ \bibnamefont {Kim}},\ }\href {\doibase
  10.1109/LED.2021.3066624} {\bibfield  {journal} {\bibinfo  {journal} {IEEE
  Electron Device Letters}\ }\textbf {\bibinfo {volume} {42}},\ \bibinfo
  {pages} {708} (\bibinfo {year} {2021})}\BibitemShut {NoStop}%
\bibitem [{\citenamefont {Kang}\ \emph {et~al.}(2018)\citenamefont {Kang},
  \citenamefont {Kim}, \citenamefont {Chung}, \citenamefont {Lee},\ and\
  \citenamefont {Kim}}]{Kang2018}%
  \BibitemOpen
  \bibfield  {author} {\bibinfo {author} {\bibfnamefont {B.~H.}\ \bibnamefont
  {Kang}}, \bibinfo {author} {\bibfnamefont {W.-G.}\ \bibnamefont {Kim}},
  \bibinfo {author} {\bibfnamefont {J.}~\bibnamefont {Chung}}, \bibinfo
  {author} {\bibfnamefont {J.~H.}\ \bibnamefont {Lee}}, \ and\ \bibinfo
  {author} {\bibfnamefont {H.~J.}\ \bibnamefont {Kim}},\ }\href {\doibase
  10.1021/acsami.7b17897} {\bibfield  {journal} {\bibinfo  {journal} {ACS
  Applied Materials and Interfaces}\ }\textbf {\bibinfo {volume} {10}},\
  \bibinfo {pages} {7223} (\bibinfo {year} {2018})}\BibitemShut {NoStop}%
\bibitem [{\citenamefont {Rajachidambaram}\ \emph {et~al.}(2012)\citenamefont
  {Rajachidambaram}, \citenamefont {Sanghavi}, \citenamefont {Nachimuthu},
  \citenamefont {Shutthanandan}, \citenamefont {Varga}, \citenamefont {Flynn},
  \citenamefont {Thevuthasan},\ and\ \citenamefont
  {Herman}}]{Rajachidambaram2012}%
  \BibitemOpen
  \bibfield  {author} {\bibinfo {author} {\bibfnamefont {J.~S.}\ \bibnamefont
  {Rajachidambaram}}, \bibinfo {author} {\bibfnamefont {S.}~\bibnamefont
  {Sanghavi}}, \bibinfo {author} {\bibfnamefont {P.}~\bibnamefont
  {Nachimuthu}}, \bibinfo {author} {\bibfnamefont {V.}~\bibnamefont
  {Shutthanandan}}, \bibinfo {author} {\bibfnamefont {T.}~\bibnamefont
  {Varga}}, \bibinfo {author} {\bibfnamefont {B.}~\bibnamefont {Flynn}},
  \bibinfo {author} {\bibfnamefont {S.}~\bibnamefont {Thevuthasan}}, \ and\
  \bibinfo {author} {\bibfnamefont {G.~S.}\ \bibnamefont {Herman}},\ }\href
  {\doibase 10.1557/jmr.2012.170} {\bibfield  {journal} {\bibinfo  {journal}
  {Journal of Materials Research}\ }\textbf {\bibinfo {volume} {27}},\ \bibinfo
  {pages} {2309} (\bibinfo {year} {2012})}\BibitemShut {NoStop}%
\bibitem [{\citenamefont {Du}\ \emph {et~al.}(2014)\citenamefont {Du},
  \citenamefont {Flynn}, \citenamefont {Motley}, \citenamefont {Stickle},
  \citenamefont {Bluhm},\ and\ \citenamefont {Herman}}]{Du2014}%
  \BibitemOpen
  \bibfield  {author} {\bibinfo {author} {\bibfnamefont {X.}~\bibnamefont
  {Du}}, \bibinfo {author} {\bibfnamefont {B.~T.}\ \bibnamefont {Flynn}},
  \bibinfo {author} {\bibfnamefont {J.~R.}\ \bibnamefont {Motley}}, \bibinfo
  {author} {\bibfnamefont {W.~F.}\ \bibnamefont {Stickle}}, \bibinfo {author}
  {\bibfnamefont {H.}~\bibnamefont {Bluhm}}, \ and\ \bibinfo {author}
  {\bibfnamefont {G.~S.}\ \bibnamefont {Herman}},\ }\href {\doibase
  10.1149/2.010409jss} {\bibfield  {journal} {\bibinfo  {journal} {ECS Journal
  of Solid State Science and Technology}\ }\textbf {\bibinfo {volume} {3}},\
  \bibinfo {pages} {Q3045} (\bibinfo {year} {2014})}\BibitemShut {NoStop}%
\bibitem [{\citenamefont {Urbach}(1953)}]{Urbach1953}%
  \BibitemOpen
  \bibfield  {author} {\bibinfo {author} {\bibfnamefont {F.}~\bibnamefont
  {Urbach}},\ }\href {\doibase 10.1103/PhysRev.92.1324} {\bibfield  {journal}
  {\bibinfo  {journal} {Physical Review}\ }\textbf {\bibinfo {volume} {92}},\
  \bibinfo {pages} {1324} (\bibinfo {year} {1953})}\BibitemShut {NoStop}%
\bibitem [{\citenamefont {Jang}\ \emph {et~al.}(2015)\citenamefont {Jang},
  \citenamefont {Park}, \citenamefont {Ahn}, \citenamefont {Kim}, \citenamefont
  {Choi}, \citenamefont {Kim},\ and\ \citenamefont {Kim}}]{Jang2015}%
  \BibitemOpen
  \bibfield  {author} {\bibinfo {author} {\bibfnamefont {J.~T.}\ \bibnamefont
  {Jang}}, \bibinfo {author} {\bibfnamefont {J.}~\bibnamefont {Park}}, \bibinfo
  {author} {\bibfnamefont {B.~D.}\ \bibnamefont {Ahn}}, \bibinfo {author}
  {\bibfnamefont {D.~M.}\ \bibnamefont {Kim}}, \bibinfo {author} {\bibfnamefont
  {S.~J.}\ \bibnamefont {Choi}}, \bibinfo {author} {\bibfnamefont {H.~S.}\
  \bibnamefont {Kim}}, \ and\ \bibinfo {author} {\bibfnamefont {D.~H.}\
  \bibnamefont {Kim}},\ }\href {\doibase 10.1021/acsami.5b04152} {\bibfield
  {journal} {\bibinfo  {journal} {ACS Applied Materials and Interfaces}\
  }\textbf {\bibinfo {volume} {7}},\ \bibinfo {pages} {15570} (\bibinfo {year}
  {2015})}\BibitemShut {NoStop}%
\bibitem [{\citenamefont {Jia}\ \emph {et~al.}(2018)\citenamefont {Jia},
  \citenamefont {Suko}, \citenamefont {Shigesato}, \citenamefont {Okajima},
  \citenamefont {Inoue},\ and\ \citenamefont {Hosomi}}]{Jia2018}%
  \BibitemOpen
  \bibfield  {author} {\bibinfo {author} {\bibfnamefont {J.}~\bibnamefont
  {Jia}}, \bibinfo {author} {\bibfnamefont {A.}~\bibnamefont {Suko}}, \bibinfo
  {author} {\bibfnamefont {Y.}~\bibnamefont {Shigesato}}, \bibinfo {author}
  {\bibfnamefont {T.}~\bibnamefont {Okajima}}, \bibinfo {author} {\bibfnamefont
  {K.}~\bibnamefont {Inoue}}, \ and\ \bibinfo {author} {\bibfnamefont
  {H.}~\bibnamefont {Hosomi}},\ }\href {\doibase
  10.1103/PhysRevApplied.9.014018} {\bibfield  {journal} {\bibinfo  {journal}
  {Physical Review Applied}\ }\textbf {\bibinfo {volume} {9}},\ \bibinfo
  {pages} {14018} (\bibinfo {year} {2018})}\BibitemShut {NoStop}%
\bibitem [{\citenamefont {de~Walle}\ and\ \citenamefont
  {Neugebauer}(2003)}]{VanDeWalle2003}%
  \BibitemOpen
  \bibfield  {author} {\bibinfo {author} {\bibfnamefont {C.~G.~V.}\
  \bibnamefont {de~Walle}}\ and\ \bibinfo {author} {\bibfnamefont
  {J.}~\bibnamefont {Neugebauer}},\ }\href {\doibase 10.1038/nature01665}
  {\bibfield  {journal} {\bibinfo  {journal} {Nature}\ }\textbf {\bibinfo
  {volume} {423}},\ \bibinfo {pages} {626} (\bibinfo {year}
  {2003})}\BibitemShut {NoStop}%
\bibitem [{\citenamefont {de~Walle}\ and\ \citenamefont
  {Neugebauer}(2006)}]{VanDeWalle2006Review}%
  \BibitemOpen
  \bibfield  {author} {\bibinfo {author} {\bibfnamefont {C.~G.~V.}\
  \bibnamefont {de~Walle}}\ and\ \bibinfo {author} {\bibfnamefont
  {J.}~\bibnamefont {Neugebauer}},\ }\href {\doibase
  10.1146/annurev.matsci.36.010705.155428} {\bibfield  {journal} {\bibinfo
  {journal} {Annual Review of Materials Research}\ }\textbf {\bibinfo {volume}
  {36}},\ \bibinfo {pages} {179} (\bibinfo {year} {2006})}\BibitemShut
  {NoStop}%
\end{thebibliography}%
\bibliographystyle{aipnum4-1}   %>>>> makes bibtex use spiebib.bst

%\textit{Previous Version Conclusions:} Hydrogen acts as a negative-U defect in an a-IGZO TFT, forming a $\mathrm{{[{O_{O}^{2-}}{H^+}]}^{1-}}$ state with VB oxygen atoms that results in a negative shift in the device $\mathrm{I_{D} - V_{G}}$ transfer curve turn-on voltage and enhanced photoconduction near the valence band. Ultrabroadband photoconduction spectra indicate that hydrogen incorporation enhances the response photoconduction response from $\mathrm{{[{O_{O}^{2-}}{H^+}]}^{1-}}$-related peak centered at around $2.8$ eV below the conduction band in a top-gate a-IGZO TFT; this effect increases as a function of H incorporation into the device, as deduced from the shift in $\mathrm{I_{D} - V_{G}}$ curve $\mathrm{V_{ON}}$. H incorporation also modifies the transient decay of the photoconduction signal of a top-gate a-IGZO TFT, suppressing the free carrier recombination time at near-VB photoexcitation energies due to the increase in the $\mathrm{{[{O_{O}^{2-}}{H^+}]}^{1-}}$ state density. The observed link between a-IGZO device electrical characteristics and deep-level subgap states modified by H incorporation motivates future UBPC-based work into the nature of the a-IGZO subgap.

\end{document}